\documentclass[aps,prd,11pt,nofootinbib,tightenlines,preprintnumbers,superscriptaddress,notitlepage]{revtex4-2}
\usepackage[a4paper,left=2.5cm,right=2.5cm, top=2.0cm,bottom=2.0cm]{geometry}
\usepackage[english]{babel}
\usepackage[applemac]{inputenc}
\usepackage[colorlinks,citecolor=blue,linktoc=all,linkcolor=black,urlcolor=black]{hyperref}
\usepackage{amsfonts,amsmath,amssymb}
\usepackage{graphicx}
\usepackage{booktabs, multirow}
\usepackage[detect-all]{siunitx}
\usepackage{hyperxmp}
\usepackage{color}
    
\newcommand{\Fig}[1]{Fig.~\ref{#1}}
\newcommand{\Eq}[1]{Eq.~(\ref{#1})}
\newcommand{\Section}[1]{Section~\ref{#1}}
\newcommand{\Table}[1]{Table~\ref{#1}}

\newcommand{\phys}{\mathrm{phys}}
\newcommand{\syst}{\mathrm{syst}}
\newcommand{\stat}{\mathrm{stat}}
\newcommand{\dof}{\mathrm{d.o.f.}}
\newcommand{\fm}{\mathrm{fm}}
\newcommand{\conn}{\mathrm{conn}}
\newcommand{\disc}{\mathrm{disc}}
\newcommand{\MeV}{\mathrm{MeV}}
\newcommand{\GeV}{\mathrm{GeV}}
\newcommand{\FF}{{\cal F}_{P\gamma^*\gamma^*}}
\newcommand{\etap}{\eta^{\prime}}
\newcommand{\ahlbl}{a_\mu^{\rm hlbl}}

\newcommand{\chib}{\overline{\chi}}
\newcommand{\Tr}{\mathrm{Tr}}
\newcommand{\dd}{\mathrm{d}}
\newcommand{\MS}{$\overline{\mathrm{MS}}$}
\newcommand{\Lb}{\overline{\mathcal{L}}}
\newcommand{\Lsym}{\mathcal{L}^{\rm sym}_{[\rho,\sigma];\mu\nu\lambda}}
\newcommand{\Llda}{\Lb^{(\Lambda)}_{[\rho,\sigma];\mu\nu\lambda}}
	
\newcommand{\Rcut}{R_{\rm cut}}	
\newcommand{\ycut}{y_{\rm cut}}

\makeatother

\begin{document}

\title{Hadronic light-by-light scattering contribution to the anomalous magnetic moment of the muon at the physical pion mass}

\author{Zoltan Fodor} 
\affiliation{Department of Physics, University of Wuppertal, D-42119 Wuppertal, Germany} 
\affiliation{J\"ulich Supercomputing Centre, Forschungszentrum J\"ulich, D-52428 J\"ulich, Germany} 
\affiliation{Institute for Theoretical Physics, E\"otv\"os University, H-1117 Budapest, Hungary} 
\affiliation{Physics Department, Pennsylvania State University, University Park, PA 16802, USA} 
\author{Antoine G\'erardin}
\email{antoine.gerardin@cpt.univ-mrs.fr}
\affiliation{Aix-Marseille Universit\'e, Universit\'e de Toulon, CNRS, CPT, Marseille, France}
\author{Laurent Lellouch} 
\affiliation{Aix-Marseille Universit\'e, Universit\'e de Toulon, CNRS, CPT, Marseille, France}
\author{Kalman K. Szabo}
\affiliation{J\"ulich Supercomputing Centre, Forschungszentrum J\"ulich, D-52428 J\"ulich, Germany}
\author{Balint C. Toth}
\affiliation{Department of Physics, University of Wuppertal, D-42119 Wuppertal, Germany}
\affiliation{J\"ulich Supercomputing Centre, Forschungszentrum J\"ulich, D-52428 J\"ulich, Germany}
\author{Christian Zimmermann}
\email{christian.zimmermann@univ-amu.fr}
\affiliation{Aix-Marseille Universit\'e, Universit\'e de Toulon, CNRS, CPT, Marseille, France}

\begin{abstract}
We present a lattice QCD calculation of the hadronic light-by-light scattering contribution to the anomalous magnetic moment of the muon using $N_f=2+1+1$ flavors of  staggered quarks with masses tuned to their physical values. Our final result, in the continuum limit, reads 
$a_{\mu}^{\rm hlbl} = 125.5(11.6)_{\stat}(0.4)_{\syst} \times 10^{-11}$ 
where the first error is statistical and the second is systematic. Light, strange and charm-quark contributions are considered. In addition to the connected and leading disconnected contributions, we also include an estimate of the sub-leading disconnected diagrams. Our result is compatible with previous lattice QCD and  data-driven dispersive determinations.
\end{abstract}

\maketitle

\section{Introduction \label{sec:introduction} }

\begin{figure}[t]
	\includegraphics*[width=0.21\linewidth]{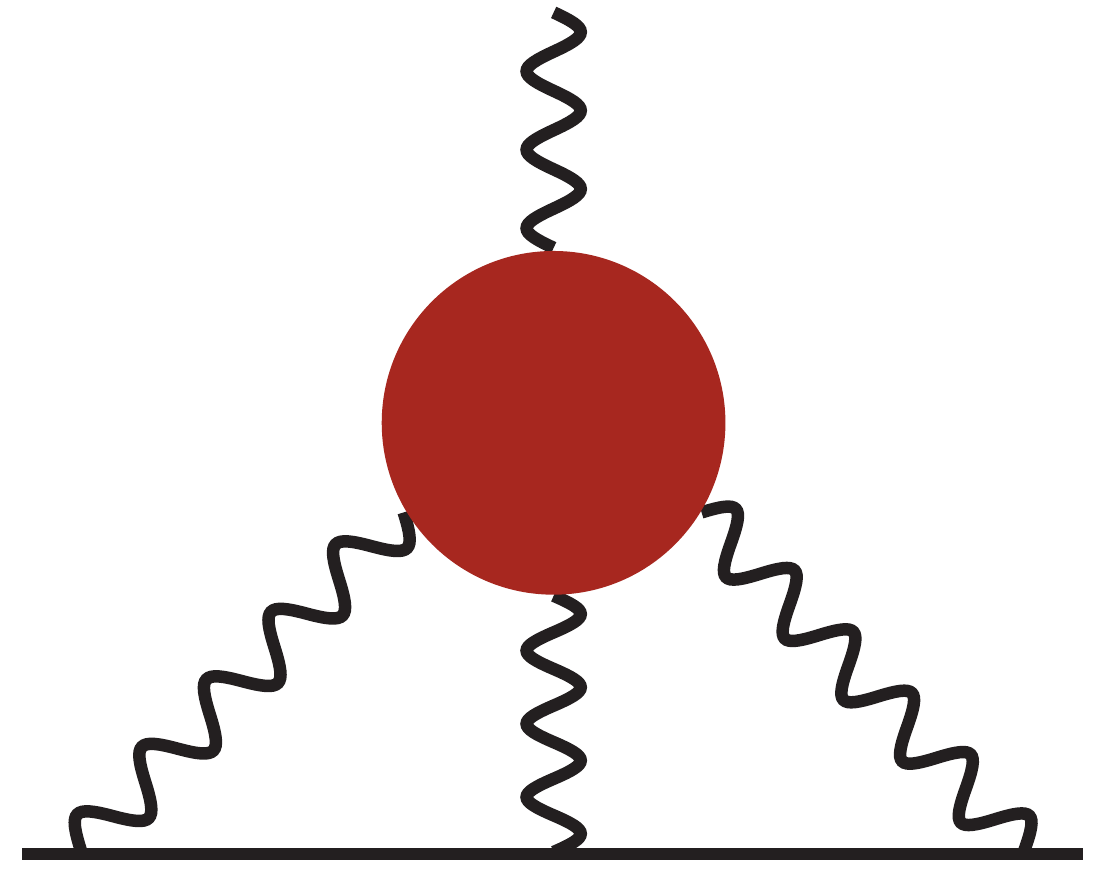}
	\caption{The HLbL contribution to the muon $g-2$. The plain and wavy lines represent the muon and the photons, respectively. The red blob represents the non-perturbative hadronic interactions. }
\label{fig:hlbl}
\end{figure}

The anomalous magnetic moment of the muon, the deviation of the gyromagnetic ratio $g_\mu$ from the value predicted by the Dirac equation, is one of the most precisely known quantities in particle physics. In 2023, this observable was measured with a precision of 0.21 ppm by the E989 Fermilab experiment \cite{Muong-2:2021ojo,Muong-2:2023cdq}, confirming the previous value obtained at Brookhaven~\cite{Muong-2:2006rrc}. The Fermilab experiment plans to reduce the uncertainties down to below 0.14 ppm while a second experiment is in preparation at J-PARC~\cite{Abe:2019thb}. 
On the theory side, the uncertainty of the Standard Model (SM) estimate quoted by the $g-2$ Theory Initiative~\cite{Aoyama:2020ynm} is about twice as large as  the current experimental uncertainty and presents a discrepancy of 5 standard deviations, although this tension is strongly suppressed by more recent lattice calculations~\cite{Aubin:2019usy,Ce:2022kxy,ExtendedTwistedMass:2022jpw,FermilabLatticeHPQCD:2023jof,RBC:2023pvn,Boccaletti:2024guq,Djukanovic:2024cmq}.

The precision of the SM prediction is limited by two hadronic processes: the hadronic vacuum polarization (LO-HVP) and the hadronic light-by-light scattering (HLbL) contribution, depicted in~\Fig{fig:hlbl}. 
Although the HLbL contribution is suppressed by an additional factor of the electromagnetic coupling, compared to the LO-HVP, its contribution to the total error budget is not negligible. A relative precision below 10\% is needed to match the future experimental precision.

A first-principle determination with a controlled error budget is a challenging task. In recent years, two different approaches have been proposed. 
First, the data-driven approach, where the HLbL diagram is calculated using dispersion relations~\cite{Melnikov:2003xd,Masjuan:2017tvw,Colangelo:2017fiz,Hoferichter:2018kwz,Gerardin:2019vio,Bijnens:2019ghy,Colangelo:2019uex,Pauk:2014rta,Danilkin:2016hnh,Jegerlehner:2017gek,Knecht:2018sci,Eichmann:2019bqf,Roig:2019reh}. This effort lead to the 2020 white paper estimate 
$\ahlbl = 92(19) \times 10^{-11}$~\cite{Aoyama:2020ynm}. 
In this framework, the dominant contribution is given by the light-pseudoscalar pole contribution for which the non-perturbative dynamics is encoded in the pseudoscalar transition form factors for space-like virtualities, a quantity accessible with lattice methods. 
Several lattice collaborations have presented results for the pion-pole contribution~\cite{Gerardin:2016cqj,Gerardin:2019vio,ExtendedTwistedMass:2023hin,Gerardin:2023naa,Lin:2024khg}. Recently, in~\cite{Gerardin:2023naa}, we presented the first calculation of the $\pi^0$, $\eta$ and $\etap$ contributions leading to the light-pseudoscalar contribution $a_{\mu}^{\rm hlbl;ps} = 85.1(5.2) \times 10^{-11}$. 
In this paper, we focus on the second approach: a direct lattice  QCD calculation of the HLbL diagram. 
Two collaborations, RBC/UKQCD~\cite{Blum:2023vlm,Blum:2017cer} and the Mainz group~\cite{Chao:2022xzg,Chao:2021tvp,Chao:2020kwq} have presented complete results so far. Both treat the QED part of the diagram, shown in~\Fig{fig:hlbl}, in the continuum and infinite volume limit while the four-point function is evaluated on the lattice. We note that a treatment of QED in finite volume has also been considered by the RBC/UKQCD collaboration in~\cite{Blum:2019ugy}.

In this paper, we follow the strategy proposed by the Mainz group~\cite{Asmussen:2022oql}, adapted to the staggered-quark discretization, and with gauge ensembles generated directly at the physical pion mass. For the dominant connected and leading quark-disconnected diagrams, three lattice spacings are used to extrapolate the light-quark contribution to the continuum limit. 
The knowledge of the light-pseudoscalar transition form factors is used to improve the estimate of the long-distance contribution and to correct our data for finite-volume effects. 
Continuum results for both connected and quark-disconnected contributions are provided to allow for cross-checks with existing and future determinations.  
Finally, we also present results for the strange and charm quark contributions as well as for all sub-leading quark-disconnected contributions that vanish in the SU(3) flavor-symmetric limit. 

The paper is organized as follows. In \Section{sec:methodology} we present the methodology for calculating $\ahlbl$ using a position space approach where the QED part of the diagram is evaluated in the continuum and in infinite volume. We present the adaptations required by the use of staggered quarks and we provide a first test of the method with a lattice calculation of the lepton-loop contribution to light-by-light scattering. In \Section{sec:light}, we focus on the dominant connected and leading disconnected light-quark contributions. In Sections~\ref{sec:strange} and~\ref{sec:charm} we present our results for the strange and charm-quark contributions respectively. Finally, in \Section{sec:sub}, we discuss the sub-leading quark-disconnected contributions before concluding in \Section{sec:ccl}.

\section{Methodology \label{sec:methodology} }

\subsection{Position-space master formula}

Our work is based on the position-space approach developed by the Mainz group in~\cite{Asmussen:2022oql} where the hadronic light-by-light scattering contribution to the muon $(g-2)$ is written as a convolution of a weight function, which describes the QED part of the diagrams depicted in~\Fig{fig:hlbl}, with a hadronic four-point correlation (represented by the red blob) computed on the lattice. In the continuum and infinite volume, the master formula reads
\begin{equation}
\ahlbl = - \frac{m_{\mu} e^6}{3} \int \dd^4y \int \dd^4x  \, \Lsym(x,y)  \, \int \dd^4z \, z_{\rho} \, \Pi_{\mu\nu\sigma\lambda}(x,y,z) \,,
\label{eq:master}
\end{equation}
with $\Lsym(x,y)$ the QED weight function that is defined below, and
\begin{equation}
\Pi_{\mu\nu\sigma\lambda}(x,y,z) = \langle  j_{\mu}(x) j_{\nu}(y) j_{\sigma}(z) j_{\lambda}(0) \rangle
\label{eq:Pihat}
\end{equation}
the four-point correlation function of the hadronic part of the electromagnetic current 
\begin{equation}
j_{\mu}(x) = \frac{2}{3} (\overline{u}\gamma_{\mu}u)(x) - \frac{1}{3} (\overline{d}\gamma_{\mu}d)(x) - \frac{1}{3} (\overline{s}\gamma_{\mu}s)(x) + \frac{2}{3} (\overline{c}\gamma_{\mu}c)(x) = \sum_f \mathcal{Q}_f \, (\overline{q}_f \gamma_\mu q_f) \,.
\label{eq:emcur}
\end{equation}

In the continuum and in infinite volume, as a consequence of current conservation, the QED weight function is not unique~\cite{Chao:2020kwq,Blum:2017cer}. 
In~\cite{Asmussen:2022oql}, the authors introduced the subtracted kernel 
\begin{align}\label{eq:lamsub}
\Llda(x,y) &= \Lb_{[\rho,\sigma];\mu\nu\lambda}(x,y)\nonumber \\ 
&-\partial_\mu^{(x)} (x_\alpha e^{-\Lambda m_{\mu}^2 x^2/2}) \Lb_{[\rho,\sigma];\alpha\nu\lambda}(0,y) - \partial_\nu^{(y)} (y_\alpha e^{-\Lambda m_{\mu}^2 y^2/2})\Lb_{[\rho,\sigma];\mu\alpha\lambda}(x,0) \,,
\end{align}
with the choice $\Lambda = 0.4$. This subtraction is a compromise between improving the behavior of the lattice data at short distance and reducing long-distance effects~\cite{Chao:2020kwq}. 
This specific weight function is the one used in the lattice calculations presented in~\cite{Chao:2020kwq,Chao:2021tvp,Chao:2022xzg}. 
The hadronic four-point function $\Pi_{\mu\nu\lambda\sigma}(x,y,z)$ satisfies the Bose symmetries
\begin{align}
\Pi_{\mu\nu\sigma\lambda}(x,y,z) = \Pi_{\nu\mu\sigma\lambda}(y,x,z) = \Pi_{\lambda\nu\sigma\mu}(-x,y-x,z-x) \,,
\end{align}
which can be imposed to the weight function. Doing so, we obtain the maximally symmetric kernel\footnote{We note that the subtraction and the symmetrization of the kernel do not commute in general.}  introduced in~\Eq{eq:master}
\label{eq:kersym}
\begin{align} 
\nonumber \Lsym(x,y) = & \frac{1}{6} \left(  
\Lb^{(\Lambda)}_{[\rho,\sigma],\mu\nu\lambda}(x,y) +
\Lb^{(\Lambda)}_{[\rho,\sigma],\nu\mu\lambda}(y,x) - 
\Lb^{(\Lambda)}_{[\rho,\sigma],\lambda\nu\mu}(x,x-y) \right. \\
& \left. - \Lb^{(\Lambda)}_{[\rho,\sigma],\nu\lambda\mu}(x-y,x) -  
\Lb^{(\Lambda)}_{[\rho,\sigma],\lambda\mu\nu}(y,y-x) - 
\Lb^{(\Lambda)}_{[\rho,\sigma],\mu\lambda\nu}(y-x,y) \right) \,.
\end{align}
This specific choice has several advantages. First, the role of the $x$ and $y$ variables becomes symmetric. Since we always perform the $z$-integral explicitly on the lattice, this means that the integrands as a function of $x$ or $y$ are the same. 
In addition, the numerical implementation allows for specific optimizations that are important to make the cost associated with the evaluation of the weight function completely sub-dominant. 

In practice, both sums over the $z$ and $x$ vertices are performed explicitly on the lattice. 
Since the resulting quantity has no open Lorentz index, it is a Lorentz scalar and, therefore, only depends on the invariant distance $y^2$. Hence, in order to perform the remaining integral over $y$, it is sufficient to sample the integrand for a few values of $|y|$ with $y/a = m \vec{n}$ and $m \in \mathbb{Z}$ with a weight given by the 3d-volume of the 3-sphere, $2\pi^2|y|^3$. In practice, we use either $\vec{n} = (1,1,1,1)$ or $\vec{n} = (3,1,1,1)$ depending on the aspect ratio $T/L$ of the lattice. 
We define the partial sum
\begin{equation}
a_{\mu}(|y|) = \int_0^{|y|} \mathcal{I}(|y^{\prime}|) \, \dd |y^{\prime}| \,,
\label{eq:amuy}
\end{equation}
where the trapezoidal rule is used to evaluate the integral.
We emphasize that the shape of the integrand depends on the specific choice of the QED weight function, although the integral does not. In particular, this prevents us  from a direct comparison of the integrand with other collaborations. 

\begin{figure}[t]
	\includegraphics*[width=0.65\linewidth]{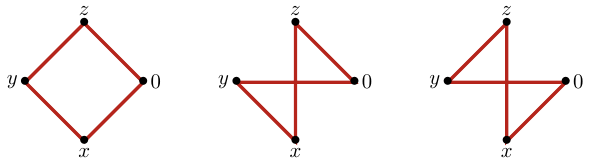}
	\caption{The three connected Wick contractions appearing in the calculation of the four-point function in~\Eq{eq:Pihat}. They are respectively denoted $\Pi^{\conn,(1)}_{\mu\nu\sigma\lambda}$, $\Pi^{\conn,(2)}_{\mu\nu\sigma\lambda}$ and $\Pi^{\conn,(3)}_{\mu\nu\sigma\lambda}$ from left to right. Each diagram appears twice with opposite fermionic flow.}
\label{fig:conn}
\end{figure}

\begin{figure}[t]
	\includegraphics*[width=0.7\linewidth]{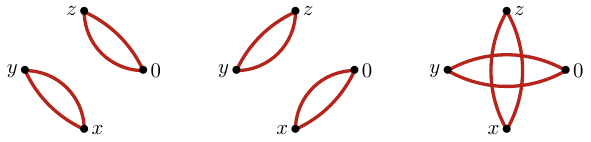}
	\caption{The three Wick contractions for the leading disconnected contribution. They are respectively denoted $\Pi^{\disc,(1)}_{\mu\nu\sigma\lambda}$, $\Pi^{\disc,(2)}_{\mu\nu\sigma\lambda}$ and $\Pi^{\disc,(3)}_{\mu\nu\sigma\lambda}$ from left to right.}
\label{fig:disc}
\end{figure}

The calculation of the hadronic four-point function~(\ref{eq:Pihat}) can be decomposed into five classes of diagrams: the connected contribution (\Fig{fig:conn}),
the leading (2+2) quark-disconnected contribution (\Fig{fig:disc}) and three classes of sub-leading disconnected diagrams that are classified according to the number of closed vector loops. The latter are not considered in this section and they are discussed separately in~\Section{sec:sub}. 

For the connected contribution, the Wick contractions are listed in \Fig{fig:conn}. As noted in~\cite{Chao:2020kwq}, evaluating all diagrams is numerically expensive as the second and third ones would require sequential inversions if one wants to sum over $x$ and $z$ explicitly. Using translational invariance in infinite volume, we obtain the relations
\begin{subequations}
\label{eq:Pi_sym}
\begin{align}
\Pi^{\conn,(2)}_{\mu\nu\sigma\lambda}(x,y,z) &= \Pi^{\conn,(1)}_{\nu\mu\sigma\lambda}(y,x,z) \,, \\
\Pi^{\conn,(3)}_{\mu\nu\sigma\lambda}(x,y,z) &= \Pi^{\conn,(1)}_{\lambda\nu\sigma\mu}(-x,y-x,z-x) \,,
\end{align}
\end{subequations}
with
\begin{equation}
\Pi^{\conn,(f),(1)}_{\mu\nu\sigma\lambda}(x,y,z) = -2 \mathrm{Re} \ \langle \Tr\left[ G_f (0,x) \gamma_{\mu} G_f (x,y) \gamma_{\nu} G_f (y,z) \gamma_{\sigma} G_f (z,0) \gamma_{\lambda} \right] \rangle_U
\end{equation}
where $G_f$ denotes the quark propagator with flavor $f$. Thus, the second and third Wick contractions can be expressed in terms of the first one, leading to our final lattice estimator
\begin{equation}
\mathcal{I}^{\conn}(|y|) = \frac{m_\mu e^6}{3} \sum_{f} \mathcal{Q}_f^4 \, 2\pi^2 |y|^3 \sum_{x,z} \Lsym(x,y)\ (x_\rho - 3z_\rho) \ \Pi^{\conn,(f),(1)}_{\mu\nu\sigma\lambda}(x,y,z) \,.
\label{eq:conn}
\end{equation}

We now turn to the leading quark-disconnected contribution. The list of Wick contractions is given in \Fig{fig:disc}. The main objects to compute on the lattice are the subtracted two-point functions defined as
\begin{subequations}
\label{eq:pihat}
\begin{align}
\widehat{\Pi}^{(f)}_{\mu\nu}(x,y) &= \Pi^{(f)}_{\mu\nu}(x,y) - \langle \, \Pi^{(f)}_{\mu\nu}(x,y) \, \rangle_U \,, \\
\Pi^{(f)}_{\mu\nu}(x,y) &= - \Tr\left[ \gamma_{\mu} G_f(x,y)  \gamma_{\nu} G_f(y,x) \right] \,,
 \end{align}
\end{subequations}
where $\langle \cdots \rangle_U$ denotes the ensemble average. The contribution to the four-point function from diagrams (1) and (2) reads 
\begin{subequations}
\begin{align}
\Pi^{\disc,(f_1f_2),(1)}_{\mu\nu\sigma\lambda}(x,y,z) &= \langle \widehat{\Pi}^{(f_1)}_{\mu\nu}(x,y) \, \widehat{\Pi}^{(f_2)}_{\sigma\lambda}(z,0) \rangle_U \\
\Pi^{\disc,(f_1,f_2),(2)}_{\mu\nu\sigma\lambda}(x,y,z) &= \langle \widehat{\Pi}^{(f_1)}_{\mu\lambda}(x,0) \, \widehat{\Pi}^{(f_2)}_{\nu\sigma}(y,z) \rangle_U \,.
\end{align}
\end{subequations}
Since we are explicitly summing over $x$ and $z$, it is convenient to rewrite the third diagram in \Fig{fig:disc} in terms of either of the other two diagrams
\begin{subequations}
\begin{align}
\Pi^{\disc,(f_1f_2),(3)}_{\mu\nu\sigma\lambda}(x,y,z)  &= \Pi^{\disc,(f_1f_2),(1)}_{\nu\lambda\sigma\mu}(y-x,-x,z-x)  \,, \\
\Pi^{\disc,(f_1f_2),(3)}_{\mu\nu\sigma\lambda}(x,y,z)  &= \Pi^{\disc,(f_1f_2),(2)}_{\nu\mu\sigma\lambda}(y,x,z) \,,
\end{align}
\end{subequations}
leading to two lattice estimators
\begin{subequations}
\begin{align}
\mathcal{I}_1^{(2+2)}(|y|) &= \frac{m_\mu e^6}{3} \, \sum_{f_1,f_2} \mathcal{Q}_{f_1}^2 \, \mathcal{Q}_{f_2}^2 \, 2\pi^2 |y|^3 \sum_{x,z} \Lsym(x,y)\ (x_\rho + y_\rho - 3z_\rho)\  \Pi^{\disc,(f_1f_2),(1)}_{\mu\nu\sigma\lambda}(x,y,z) \\ 
\mathcal{I}_2^{(2+2)}(|y|) &= \frac{m_\mu e^6}{3} \, \sum_{f_1,f_2} \mathcal{Q}_{f_1}^2 \, \mathcal{Q}_{f_2}^2 \, 2\pi^2 |y|^3 \sum_{x,z} \Lsym(x,y)\ (y_\rho - 3z_\rho)\  \Pi^{\disc,(f_1,f_2),(2)}_{\mu\nu\sigma\lambda}(x,y,z) \,.
\end{align}
\end{subequations}
In practice, we observe that averaging over both estimators leads to a reduction of the statistical error and we use
\begin{align}
\mathcal{I}^{(2+2)}(|y|) = \frac{1}{2} \left( \mathcal{I}_1^{(2+2)}(|y|) + \mathcal{I}_2^{(2+2)}(|y|)\right) \,.
\end{align}
Finally, thanks to the aforementioned symmetrization, the same weight function (evaluated at the same lattice sites) appears both in the connected and in the disconnected contributions.

\subsection{Implementation with staggered quarks}

In our lattice implementation, we use the conserved vector current
\begin{equation}
j_{\mu}(x) = -\frac{1}{2} \eta_{\mu}(x) \left[ \chib(x+a\hat{\mu}) U_{\mu}^{\dag}(x) \chi(x) + \chib(x) U_{\mu}(x) \chi(x+a\hat{\mu}) \right] \,,
\label{eq:V}
\end{equation}
where $\chi$ is a single component staggered fermion field and $\eta_{\mu}$ is a phase factor (we use the conventions of \cite{Follana:2006rc}, also used in~\cite{spectro}). The gauge links $U_{\mu}(x)$ ensure that correlation functions are gauge invariant. This current is the one used in the calculation of the LO-HVP contribution in~\cite{Boccaletti:2024guq,Borsanyi:2020mff} and in the calculation of the light pseudoscalar transition form factors in~\cite{Gerardin:2023naa}. This current does not require multiplicative renormalization.

In the position-space approach presented above, the sum over the vertices $x$ and $z$ are performed explicitly over the lattice but not the sum over $y$. As a consequence, the integrand $\mathcal{I}(|y|) = 2\pi^2 |y|^3 f(|y|)$, at a given site $y$, receives contributions from 16 taste-partners: the desired taste-singlet contribution appears with a factor one while other tastes contribute with oscillating factors $(-1)^{\zeta\cdot y/a}$ where $\zeta$ is one of the fifteen non-zero vectors with components in~$\{0,1\}$.

For the LO-HVP contribution in the TMR representation~\cite{Bernecker:2011gh}, the correlation function is projected on zero-momentum such that we are left with only two contributions: the vector taste-singlet and the taste non-singlet pseudovector contribution that oscillates with a factor $(-1)^{t/a}$. This is visible at the integrand level but the un-wanted contribution vanishes like $a^2$ once the sum over Euclidean time is performed. 
In the present case, we would like to project onto the vector taste singlet. This would be the case if the sum over $y$ could be performed exactly. 
Instead, we suppress the un-wanted contribution by applying a smearing function $\mathcal{S}_\mu$ to the function $f$. Thus, we evaluate the integrand $\mathcal{I}(|y|) = 2\pi^2 |y|^3 f^{\mathcal{S}}(|y|)$ with
\begin{equation}
f^{\mathcal{S}}(y) = \prod_{\mu} \mathcal{S}^{(2)}_\mu f(y) \,,
\end{equation}
and $\mathcal{S}^{(2)}_\mu f(y) = (f(y-a\hat{\mu}) + 2f(y) + f(y+a\hat{\mu}))/(4a)$. 
We note that the simpler smearing function $\mathcal{S}^{(1)}_\mu f(y) = (f(y)+f(y+a\hat{\mu})) / (2a)$ would introduce un-wanted $O(a)$ lattice artifacts.
At first sight this method seems expensive since it requires the evaluation of the integrand for many values of $y$. However, this sum can be spread among different gauge configurations such that the sum becomes exact only after ensemble average. In practice we follow a different strategy: the smearing is done explicitly on each configuration but, for each end-point, the origin of the lattice is randomly chosen. Doing so, we find that the statistical error scales with the number of inversions, so that for a given statistical uncertainty there is virtually no additional cost.

\subsection{Lepton loop on the lattice}

\begin{figure}[t]
	\includegraphics*[width=0.48\linewidth]{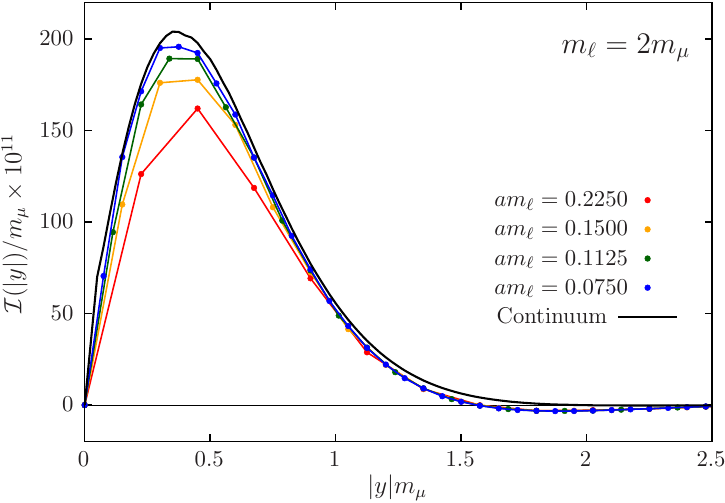}
	\includegraphics*[width=0.48\linewidth]{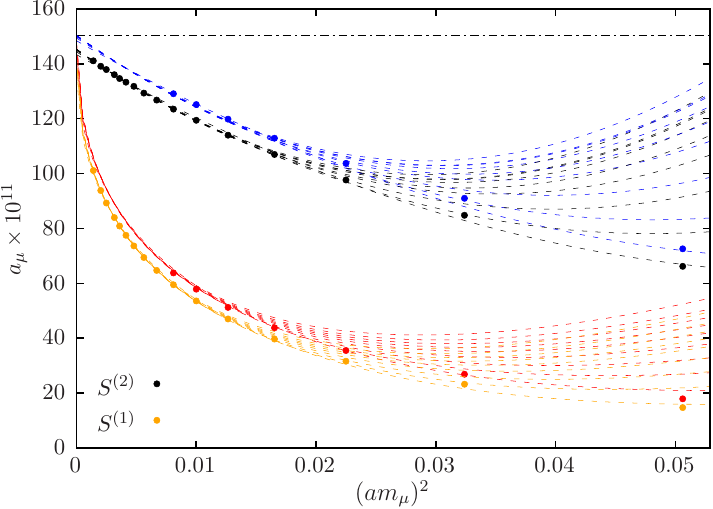}
	\caption{Left: Integrand for the lepton-loop contribution to light-by-light scattering on the lattice using $\mathcal{S}^{(2)}_\mu$. Right: Continuum extrapolations of the lepton-loop contribution for several cuts in the lattice spacing. The lepton mass is set to $m_\ell = 2 m_{\mu}$. The orange (black) dashed lines correspond to the continuum extrapolation using the smearing $\mathcal{S}^{(1)}_\mu$ ($\mathcal{S}^{(2)}_\mu$) and $L m_{\ell} = 7.2$. The red and blue curves are the finite-volume corrected extrapolations using the procedure described in the main text.}
\label{fig:lepton}
\end{figure}

Before simulating QCD, and as a first test of our setup, we have performed a lattice calculation of the lepton-loop contribution to light-by-light scattering in QED as in~\cite{Blum:2017cer,Asmussen:2022oql}. This is done by setting all SU(3) gauge links to unity, and dividing the final results by a factor of three to get rid of the sterile QCD color index. This calculation is performed using 16 lattices of size $L^4$ with $L/a \in (16,20,24,28,32,36,40,44,48,52,56,60,64,72,80,92)$, the ratio $m_\ell / m_{\mu} = 2$ and the volume $L m_{\ell} = 7.2$. The integrand for several ensembles is shown in the left panel of \Fig{fig:lepton}. Our data is extrapolated to the continuum limit assuming the functional dependence
\begin{equation}
a^{\rm lepton}_{\mu}(a) = a^{\rm lepton}_{\mu}(0) + \alpha_1 a + \alpha_2 a^2 + \alpha_4 a^4 \,,
\label{eq:extrapLL}
\end{equation}
where the parameter $\alpha_1$ is set to zero when using $S_{\mu}^{(2)}$. Results are shown in black and orange on the right panel of \Fig{fig:lepton}. 
An additional set of 7 ensembles with $L m_{\ell} = 14.4$ (blue and red points in~\Fig{fig:lepton}) shows that finite-size effects are independent of the lattice spacing. Thus, we use the difference between the large and small volumes at the finest available lattice spacing ($am_\mu = 0.09$) to estimate the finite-size correction. This leads to the blue and red curves. 
To estimate the systematic error associated to the continuum extrapolation, we have perform 8 cuts in the lattice spacing. The final value is given by the less aggressive cut (excluding the 7 coarsest lattices) and the systematic error is quoted as half the spread among all extrapolations: 
$a_{\mu}^{\mathrm{lepton};S_2} = (150.35 \pm 1.49) \times 10^{-11}$ and $a_{\mu}^{\mathrm{lepton};S_1} = (147.70 \pm 4.37) \times 10^{-11}$. 
With both smearings we are able to reproduce the known result $a^{\rm lbl}_{\mu} = 150.31 \times 10^{-11}$~\cite{Laporta:1991zw,Laporta:1992pa,Kuhn:2003pu}.

\subsection{Gauge configurations}

This work is based on a subset of ensembles presented in~\cite{Boccaletti:2024guq,Borsanyi:2020mff} and previously used in~\cite{Gerardin:2023naa}. They have been generated using $N_f = 2 + 1 + 1$ dynamical staggered fermions with four steps of stout smearing. The bare quark masses have been tuned such that the Goldstone mesons are at nearly physical pion and kaon mass.
The light quark contribution has been computed at three values of the lattice spacing using large volume ensembles with $L\approx 4$ and 6~fm. Data using smaller physical volumes are also available. The statistically more precise strange and charm quark contributions have been computed at five values of the lattice spacing, including ensembles with $L\approx 3, 4$ and 6~fm, and an additional ensemble with a finer lattice spacing, $a \approx 0.048~\fm$, is included in the analysis of the connected charm quark contribution.  The parameters and the number of gauge configurations analyzed on each ensemble are summarized in~\Table{tab:ens}.

\begin{table}[t]
\renewcommand{\arraystretch}{1.1}
\caption{Parameters of the simulations: the bare coupling $\beta = 6/g_0^2$, the lattice size, the lattice spacing $a$ and the spatial extent $L$ in physical units, the bare light and strange-quark masses and the number of gauge configurations for the light, strange and charm  contributions. An asterisk denotes an ensemble where only the connected contribution has been computed. The charm mass is set by $m_c/m_s = 11.85$~\cite{Davies:2009ih}}
\vskip 0.1in
\begin{tabular}{l@{\hskip 01em}c@{\hskip 01em}l@{\hskip 01em}c@{\hskip 01em}l@{\hskip 01em}l@{\hskip 01em}l@{\hskip 01em}l@{\hskip 01em}l@{\hskip 01em}l}
\hline
$\quad\beta\quad$	&	$L^3\times T$ 	&	$a~[\fm]$	&	$L~[\fm]$	&	$am_l$		&	$am_s$	&	$\#$confs	 &	$\#$confs	  &	$\#$confs	 & Id\\
				&				&			&			&				&	         	&	(light)	 &	(strange)	  &	(charm)	 & \\
\hline
$3.7000$	&	$48^3\times64$	& 	0.132	& 	6.3	& 0.00205349 & 0.0572911	&  		&	400		&		&	V6-L48\\
		&	$32^3\times64$	& 			& 	4.2	& 0.00205349 & 0.0572911	&  900	&	900		&	900	&	V4-L32\\
		&	$24^3\times48$	& 			&	3.2	& 0.00205349 & 0.0572911	&  700	&	700		&	700	&	V3-L24\\
\hline
$3.7553$	&	$56^3\times84$	& 	0.112	&	6.2	& 0.00171008 & 0.0476146	&  500	&	$50^{*}$	&	$50^{*}$	&	V6-L56-1\\
		&	$56^3\times84$	& 			&	6.2	& 0.00174428 & 0.0461862	&  500	&			&	$50^{*}$	&	V6-L56-2\\
		&	$28^3\times56$	& 			&	3.1	& 0.00171008 & 0.0476146	&  850	&	850		&	850		&	V6-L28\\
\hline
$3.8400$	&	$64^3\times96$	& 	0.095	&	6.1	& 0.001455     & 0.04075		&  1150	&			&	$50^{*}$	&	V6-L64\\
		&	$32^3\times64$	& 			&	3.0	& 0.00151556 & 0.0431935	&  		&	$50^{*}$	&	$50^{*}$	&	V3-L32-1\\
		&	$32^3\times64$	& 			&	3.0	& 0.00143       & 0.0431935	&   		&			&	$50^{*}$ &	V3-L32-2\\
		&	$32^3\times64$	& 			&	3.0	& 0.001455     & 0.04075		&    1000	&	1000		&	1000		&	V3-L32-3\\
		&	$32^3\times64$	& 			&	3.0	& 0.001455     & 0.03913		&   		&			&	$50^{*}$	&	V3-L32-4\\
\hline
$3.9200$	&	$40^3\times80$	& 	0.079	&	3.1	& 0.001207     & 0.032		&   		&	$200^{*}$	&	$200^{*}$	&	V3-L40-1\\
		&	$40^3\times80$	& 			&	3.1	& 0.0012         & 0.0332856	&   		&			&	$50^{*}$	&	V3-L40-2\\
\hline
$4.0126$	&	$48^3\times96$	& 	0.064	&	3.1	& 0.00095897 & 0.0264999	&   		&	$50^{*}$	&	$50^{*}$	&	V3-L48-1\\
		&	$48^3\times96$	& 			&	3.1	& 0.001002     & 0.027318		&   		&  	$50^{*}$	&	$50^{*}$	&	V3-L48-2\\
\hline
$4.1479$	&	$128^3\times192$	& 	0.048	&	6.2	& 0.000701029 & 0.0193696	&   		&			&	$25^{*}$	&	V6-L128\\
\hline
 \end{tabular} 
\label{tab:ens}
\end{table}

\section{Light-quark contribution \label{sec:light}} 

In the dispersive framework, the pion pole is the dominant contribution to $\ahlbl$~\cite{Melnikov:2003xd,Masjuan:2017tvw,Colangelo:2017fiz,Hoferichter:2018kwz,Gerardin:2019vio,Bijnens:2019ghy,Colangelo:2019uex,Pauk:2014rta,Danilkin:2016hnh,Jegerlehner:2017gek,Knecht:2018sci,Eichmann:2019bqf,Roig:2019reh}. 
Since the pion is the lightest meson, it is expected to provide a good description of the integrand at long distances but also to be responsible for the dominant  finite-size effects. Thus, before presenting our lattice data, we start this section with a discussion of the pseudoscalar-pole contribution in position space.

\subsection{The pseudoscalar-pole contribution \label{sec:ps-pole}}

In Euclidean space, the pseudoscalar-pole contribution to the fourth-rank hadronic light-by-light tensor reads~\cite{Knecht:2001qf,Asmussen:2022oql,Green:2015sra} 
\begin{align}
\widetilde{\Pi}&_{\mu\nu\sigma\lambda}(q_1,q_2,q_3)  = \nonumber \\ 
&
\frac{ \FF(-q_1^2, -q_2^2) \ \FF(-q_3^2, -(q_1+q_2+q_3)^2)}{(q_1+q_2)^2 + m_{P}^2} 
\ \varepsilon_{\mu\nu\alpha\beta} \, q_{1\alpha} q_{2\beta}
\ \varepsilon_{\sigma\lambda\gamma\delta} \, q_{3 \gamma} (q_1 + q_2)_{\delta} \nonumber \\ 
& 
+ \frac{\FF(-q_1^2, -(q_1 + q_2 + q_3)^2) \ \FF(-q_2^2, -q_3^2)}{ (q_2+q_3)^2 + m_{P}^2}  
\ \varepsilon_{\mu\lambda\alpha\beta} \, q_{1 \alpha} (q_2 + q_3)_{\beta}
\ \varepsilon_{\nu\sigma\gamma\delta} \, q_{2 \gamma} q_{3 \delta} \nonumber \\
&
+ \frac{\FF(-q_1^2, -q_3^2) \ \FF(-q_2^2, -(q_1+q_2+q_3)^2) }{ (q_1+q_3)^2 + m_{P}^2}  
\ \varepsilon_{\mu\sigma\alpha\beta} \, q_{1 \alpha} q_{3 \beta} 
\  \varepsilon_{\nu\lambda\gamma\delta} \, q_{2 \gamma} (q_1 + q_3)_{\delta} 
\label{eq:pspole}
\end{align}
with $\FF(-q_1^2, -q_2^2)$ the pseudoscalar transition form factor, evaluated for space-like virtualities and $m_P$ the pseudoscalar mass.  
For the three lightest pseudoscalars, $P=\pi^0$, $\eta$ and $\etap$, the transition form factors have been computed using staggered quarks on the set of gauge ensembles used in this work in~\cite{Gerardin:2023naa}. 

We now want to express the four-point function in position-space. The contribution from the first term in \Eq{eq:pspole} reads ($ \int_q \equiv \int \frac{ \dd^4q}{(2\pi)^4}$)
\begin{align}
\Pi_{\mu\nu\sigma\lambda}(x,y,x) &= \int_{q_1,q_2,q_3} \ \widetilde{\Pi}_{\mu\nu\sigma\lambda}(q_1,q_2,q_3) \ e^{i (q_1 x + q_2 y + q_3 z)}  \nonumber \\
&= \int_{q_1,q_2,q_3} M_{\mu\nu}(q_1,q_2) M_{\sigma\lambda}(q_3,q_1+q_2+q_3) \, G_{P}(q_1+q_2)  \ e^{i (q_1 x + q_2 y + q_3 z)}
\label{eq:Pi2}
\end{align}
with the light pseudoscalar propagator
\begin{equation}
\widetilde{G}_{P}(x) = \int_p \frac{e^{ipx}}{p^2+m_{P}^2} =  \int_p G_{P}(p) e^{ipx} 
\end{equation}
and 
\begin{equation} 
M_{\mu\nu}(q_1,q_2)  = \epsilon_{\mu\nu\alpha\beta} \, q_1^{\alpha} \, q_2^{\beta} \, \FF(-q_1^2, -q_2^2) \,.
\end{equation}
Inserting a delta function $\delta^{(4)}(q_4-q_1-q_2-q_3)$, making the replacement $u \to u+v$ followed by the replacement $v\to-v$ and introducing the notation
\begin{align}
\nonumber \widetilde{M}_{\mu\nu}(u;x,y) &= \int_{q_1,q_2}  M_{\mu\nu}(q_1,q_2) \, e^{iq_1(x-u)} e^{iq_2(y-u)} \\
&= - \varepsilon_{\mu\nu\alpha\beta}  \partial_{x_\alpha} \partial_{y_\beta} \int_{q_1,q_2} \FF(-q_1^2,-q_2^2) \, e^{iq_1(x-u)} e^{iq_2(y-u)} \,,
\label{eq:M}
\end{align}
\Eq{eq:Pi2} can be recast as
\begin{align}
\Pi_{\mu\nu\sigma\lambda}(x,y,x) &= - \int_{uv} \widetilde{M}_{\mu\nu}(u;x,y) \, \widetilde{G}_{P}(u-v) \, \widetilde{M}_{\sigma\lambda}(v;z,0) \,.
\end{align}
Similar expressions are obtained for the second and third terms in \Eq{eq:pspole} and one finally obtains the position-space formula
\begin{multline}
\Pi_{\mu\nu\sigma\lambda}(x,y,x) = - \int_{uv}  \widetilde{G}_{P}(u-v) \left( \widetilde{M}_{\mu\nu}(u;x,y) \widetilde{M}_{\sigma\lambda}(v;z,0) + \widetilde{M}_{\mu\sigma}(u;x,z) \widetilde{M}_{\nu\lambda}(v;y,0) \right. \\ \left. + \widetilde{M}_{\mu\lambda}(u;x,0) \widetilde{M}_{\nu\sigma}(v;y,z) \right) \,.
\end{multline}
This equation was already derived in~\cite{Blum:2023vlm}. 
In our QCD simulations, we take advantage of the symmetries of the four-point function in \Eq{eq:Pi_sym} to reduce the three diagrams depicted in \Fig{fig:conn} to a single diagram. 
If this procedure has no impact on the value of $\ahlbl$ in the continuum and infinite-volume limits, it affects the shape of the integrand. 
The matching between the different QCD Wick-contractions and Feynman diagrams was derived using Partially-Quenched Chiral Perturbation Theory (PQChPT) in Appendices A of~\cite{Chao:2021tvp} and~\cite{Chao:2020kwq}. With our estimator, the four-point function that needs to be evaluated in position space reads
\begin{equation}
\Pi_{\mu\nu\sigma\lambda}(x,y,x) = - \frac{3}{2} \int_{uv}  \widetilde{G}_{\pi}(u-v) \left( \widetilde{M}_{\mu\nu}(u;x,y) \widetilde{M}_{\sigma\lambda}(v;z,0) + \widetilde{M}_{\mu\lambda}(u;x,0) \widetilde{M}_{\nu\sigma}(v;y,z) \right) 
\label{eq:psnum}
\end{equation}
with $\widetilde{M}_{\mu\nu}$ defined in~\Eq{eq:M}. When the connected and quark-disconnected contributions are treated separately, we use the weight factors $\Pi^{\conn;(1)} = \frac{34}{9} \Pi$ and $\Pi^{\mathrm{(2+2)}} = -\frac{25}{9} \Pi$ for the pion contribution while we assume that the $\eta$ and $\etap$ contribute only to the disconnected diagram~\cite{Bijnens:2016hgx,Chao:2021tvp}.

\begin{figure}[t]
	\includegraphics*[width=0.49\linewidth]{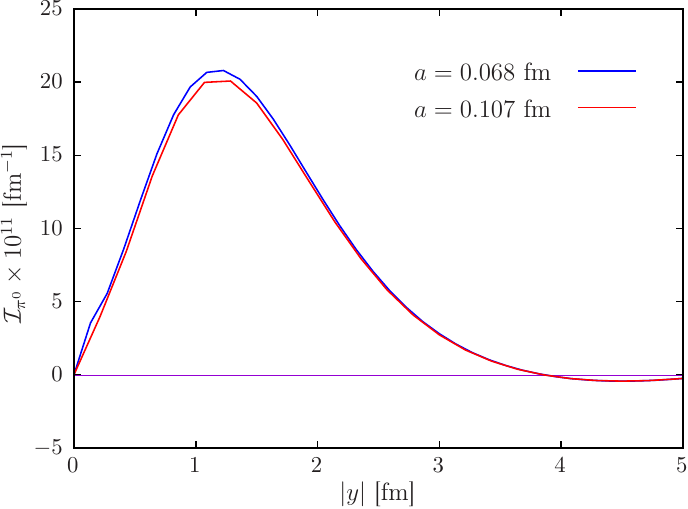}
	\includegraphics*[width=0.49\linewidth]{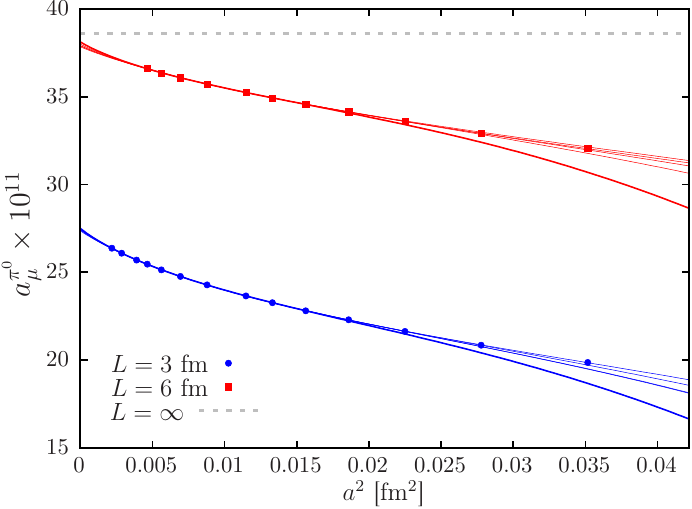}
	\caption{Left: Integrand of the pion-pole contribution at two values of the lattice spacings using an LMD parametrization. Right: Computation of the finite-volume correction for the ensembles V6-L64 and V3-L32-3 assuming an LMD parametrization of the pion transition form factor. The dashed grey line is the infinite-volume value computed in momentum-space~\cite{Knecht:2001qf}.}
\label{fig:fse}
\end{figure}

The integrand is computed in a finite-volume lattice using~\Eq{eq:conn} and with the same QED weight function. Again, both sums over $x$ and $z$ are performed explicitly and the integrand is evaluated for several values of $|y|$. 
Several lattice spacings are used to perform the continuum extrapolation (we stress that the lattice spacings discussed here are unrelated to the ones used in the QCD simulations of~\Table{tab:ens}). A typical continuum extrapolation is shown in the right panel of~\Fig{fig:fse}. For the integrand, we use the result obtained at the finest lattice spacing.  As can be seen on the left panel of~\Fig{fig:fse}, discretization effects are negligible for the tail of the integrand with $|y| > 1.5~\fm$. 

In practice, two different models are used to describe the pseudoscalar TFFs, the vector meson dominance (VMD) and lowest meson dominance (LMD) parametrization
\begin{subequations}
\label{eq:models}
\begin{align}
\FF^{\rm VMD}(Q_1^2,Q_2^2) &= \frac{ \alpha M_V^4 }{ (M_V^2 + Q_1^2) (M_V^2 + Q_2^2) }  \,, \\
\FF^{\rm LMD}(Q_1^2,Q_2^2) &= \frac{ \alpha M_V^4 - \beta (Q_1^2 + Q_2^2) }{ (M_V^2 + Q_1^2) (M_V^2 + Q_2^2) }  \,.
\end{align}
\end{subequations}
The motivation for using a VMD or LMD parametrization is to simplify the numerical evaluation of~\Eq{eq:psnum}. For example, when using the VMD model, the TFF factorizes completely in terms of the photon virtualities $Q_1^2$ and $Q_2^2$, simplifying the evaluation of \Eq{eq:M}. 
Both models provide a good description of the lattice data for moderate virtualities $Q^2 \leq 0.8~\GeV^2$ such that we expect them to provide a reasonable estimate of the finite-size correction and of the long-distance behavior of the HLbL integrand. 
However, both models differ at large virtualities. The VMD model fulfills the Brodsky-Lepage behavior~\cite{Lepage:1979zb,Lepage:1980fj,Brodsky:1981rp} in the single-virtual regime but decreases as $1/Q^4$ in the double-virtual regime while the OPE predicts a $1/Q^2$ scaling~\cite{Nesterenko:1982dn,Novikov:1983jt}. Instead, the LMD model fulfills the OPE prediction in the double-virtual regime but tends to a constant in the single-virtual regime. 
For these reasons, the VMD (LMD) tends to under(over)-estimate the transition form factors parametrized in a model-independent way in~\cite{Gerardin:2023naa} such that we expect them to provide realistic bounds.

In practice, the parameters $\alpha$, $M_V$ and $\beta$ in \Eq{eq:models} are obtained from a fit to the lattice data presented in~\cite{Gerardin:2023naa}.
For the LMD (VMD) model, the fit is restricted to the range of virtualities $Q_1^2,Q_2^2 < 1.3~\GeV^2$ ($0.8~\GeV^2$) respectively where $\chi^2/\dof \approx 1$ is obtained in all cases. 

\subsection{Data at finite lattice spacing}

\begin{figure}[t]
	\includegraphics*[width=0.49\linewidth]{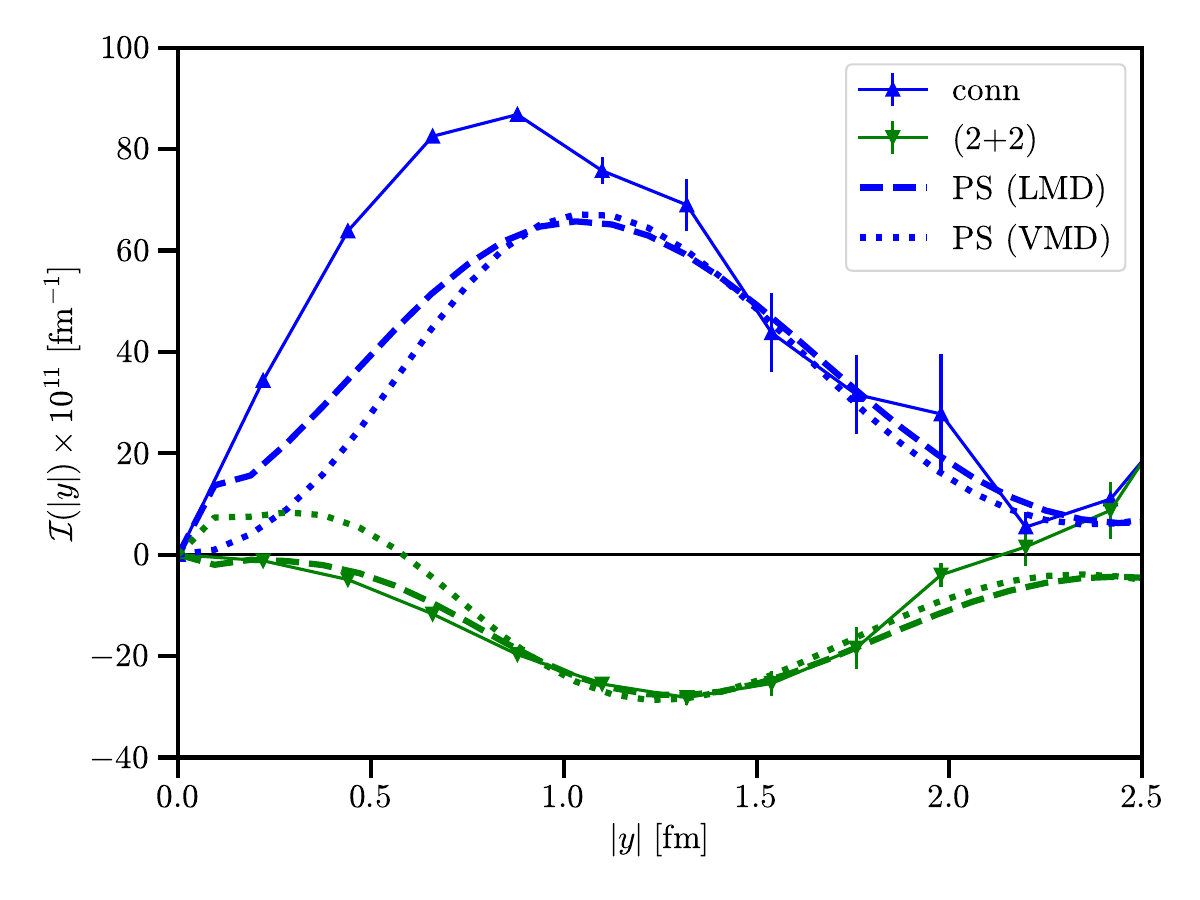}
	\includegraphics*[width=0.49\linewidth]{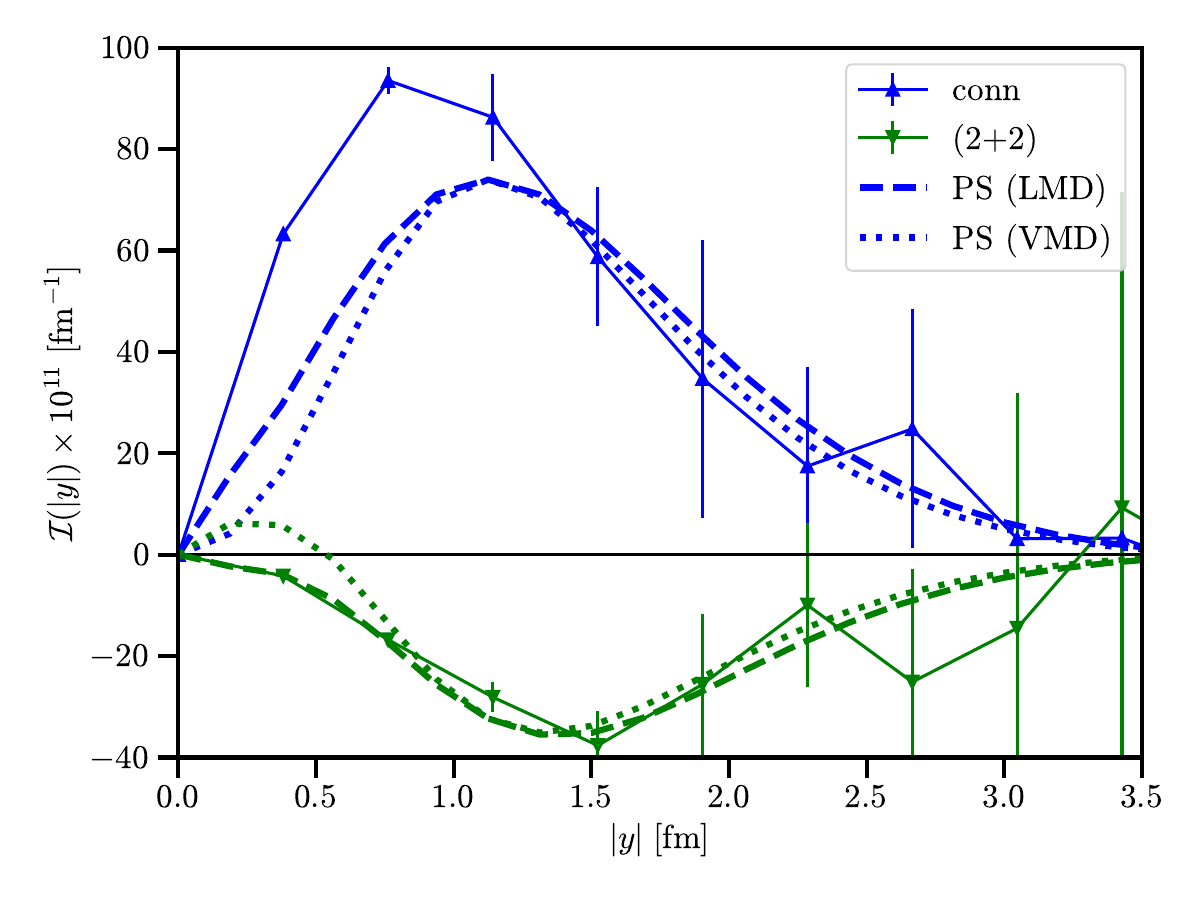}	
	\caption{Pseudoscalar-pole tail extension for two different volumes ($L\approx3~\fm$ on the left and $L \approx 6~\fm$ on the right) and using both the VMD and LMD parameterizations of the TFFs. The lattice spacing is $a \approx 0.112~\fm $.}
\label{fig:tail}
\end{figure}

In this section, we focus on the connected and leading quark-disconnected contributions. 
The integrand as a function of $|y|$ is shown in \Fig{fig:tail} for two ensembles at the same lattice spacing ($a\approx 0.112~\fm$) but with different volumes. At the physical point, the signal-to-noise ratio deteriorates rapidly and the signal is typically lost for $|y| > 2.5~\fm$ on the large-volume ensemble ($L \approx 6~\fm$). This requires a specific treatment of the tail of the integrand. 
In addition, the data needs to be corrected for finite-size effects (FSE). We now explain our methodology before presenting our results on each ensemble. 

\subsubsection{Truncation of the sum over the $x$-vertex}

\begin{figure}[t]
	\includegraphics*[width=0.49\linewidth]{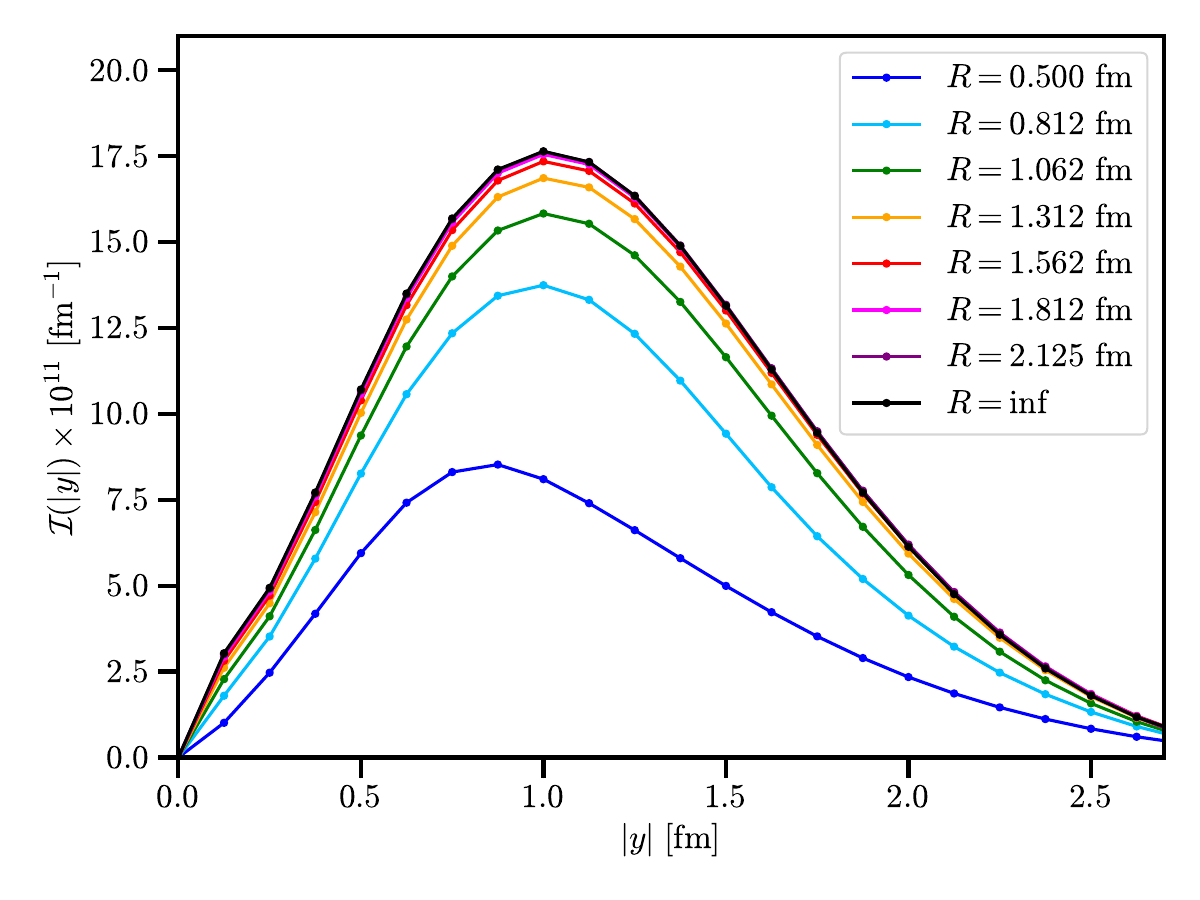}
	\includegraphics*[width=0.49\linewidth]{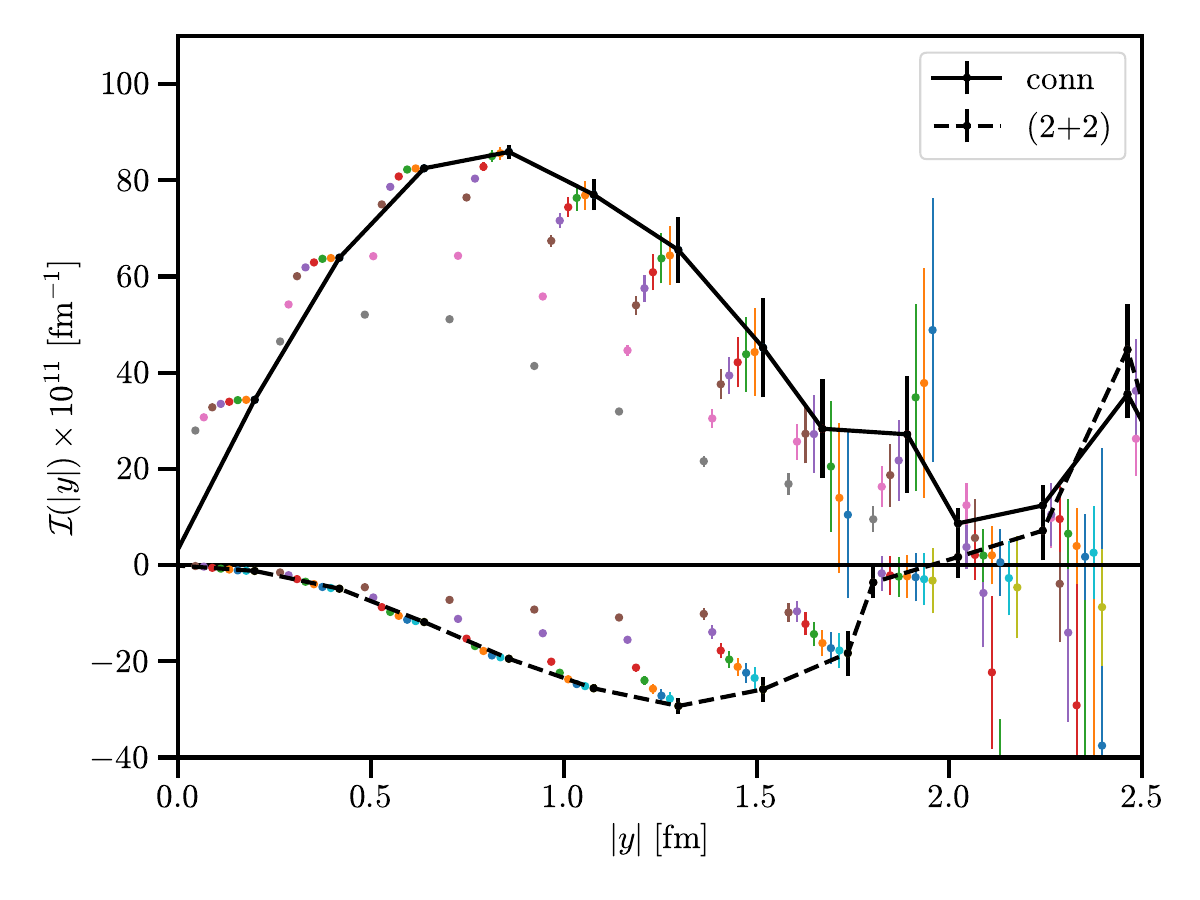}
	\caption{Left: saturation of the $x$-integration for different values of $\Rcut$ for the pion-pole contribution corresponding to our coarsest ensemble. Right: same plot for the QCD gauge ensemble V4-L28 with the cuts $\Rcut = 0.55$, 0.77, 1.10, 1.32, 1.54, 1.87, 2.09, $\infty$~fm. For a given $|y|$, the values of the integrand for the eight cuts are slightly shifted for clarity. Both the connected and leading (2+2) disconnected data are shown. The black line corresponds to our final cut selection.}
\label{fig:cuts}
\end{figure}

In~\Eq{eq:M}, the matrix element for the pion-pole contribution decreases rapidly when the vertices $x$ and $y$ are well separated. \Eq{eq:psnum} suggests that, in a region where the pion-pole contribution dominates, the bulk of the signal comes from either $x \approx y$ or $x \approx 0$ (first and second term respectively). We thus evaluate the sum over $x$ with the restriction that
$|x| < \Rcut$ or $|x-y| < \Rcut$, for several values of $\Rcut$. The results obtained for the pion-pole contribution in finite volume, using the model described in \Section{sec:ps-pole}, are shown in the left panel of~\Fig{fig:cuts}. 
We observe that more than 88\% of the signal is already captured using $R_{\rm cut} \approx 1.1~\fm$ at $|y| \geq 1.5~\fm$.
This strategy has also been tested on gauge ensembles with small physical volumes where very high statistics can be reached. Results are shown on the right panel of~\Fig{fig:cuts}. Again, we observe that the bulk of the signal comes from $\Rcut \approx 1.1~\fm$. 
In our analysis, we use the information gained on smaller volumes to select the value of $\Rcut$ on large-volume ensembles where the statistical precision is lower and where it would be difficult to select the cut with confidence without introducing a possible bias. The chosen value of $\Rcut$ depends on $|y|$ but no cut is performed for $|y| < 1.5~$fm where the statistical precision is sufficiently good.

\subsubsection{Tail of the integrand \label{eq:tail}}

In \Fig{fig:tail} we compare the tail of the integrand with the prediction from the pseudoscalar-pole contribution in position-space and in finite volume presented in~\Section{sec:ps-pole}. 
In~\cite{Bijnens:2016hgx,Jin:2016rmu,Gerardin:2017ryf}, it was shown that, for the connected diagram, the pion-pole contribution is enhanced by a factor 34/9 while the $\eta$ and $\etap$ do not contribute. Instead, in the quark-disconnected contribution, the pion-pole contributes with the factor $-25/9$ while the $\eta$ and $\etap$ poles contribute with a factor 1. Although the $\eta$ and $\etap$ contributions are small, we find that they are not negligible, in particular at coarse lattice spacings where the taste-singlet pion is not significantly lighter than the $\eta$ meson.
The reconstruction of the tail is performed using both the VMD and LMD parameterizations of the TFFs.  We observe that both models are in good agreement for $|y|>1.5~\fm$. They provide an excellent description of our lattice data both in large volume and in smaller volume where a high statistical precision is achieved. We have also checked that the uncertainties on the TFFs parameters in~\Eq{eq:models} lead to negligible effects compared to the difference between the two models and these effects are neglected here. 
In practice, our strategy is to integrate over lattice data up to some cut, $\ycut$, and then use the pseudoscalar-pole to estimate the missing tail contribution. The difference  between both parameterizations of the TFF is used as an estimate of the systematic uncertainty. 

\subsubsection{Finite-size effects}

In this work, we assume that finite-size effects are well described by the pseudoscalar-pole contribution. 
In \Eq{eq:master}, the four-point correlation function is computed using a taste-singlet vector current such that the exchanged on-shell pseudoscalar meson must be a taste-singlet. Among the sixteen pion tastes, the taste-singlet (I) pion is the most affected by taste-breaking effects and its mass varies from $m_\pi^{\rm I} \approx 430~\MeV$ at our coarsest lattice spacing, down to $m_\pi^{\rm I} \approx 290~\MeV$ at $a \approx 0.095~\fm$. Thus, we expect finite-size effects to increase as we reach the continuum limit. 
For the quark-disconnected and total light quark contributions, we include the FSE due to the $\eta$ and $\etap$ contributions. Their contribution is negligible in large volumes but not in small ones at our coarsest lattice spacing. 
For each pseudoscalar $P=\pi^0, \eta, \etap$, and each ensemble, we define the finite volume correction 
\begin{align}
v^{P} = a_\mu^{P}(\infty) - a_\mu^{P}(V) 
\end{align}
as the difference between $a_\mu^{P}(\infty)$, the pseudoscalar-pole contribution computed in infinite volume, and $a_\mu^{P}(V)$, the same quantity computed in finite volume. 
The results for all gauge ensembles are listed \Table{tab:fse2}.

\begin{table}[b]
\renewcommand{\arraystretch}{1.2}
\caption{Estimates of the finite-size-effect correction to the light-quark contribution for all ensembles, using the method described in the text. Results are given in units of $10^{-11}$. The quoted error is the systematic error from the continuum extrapolation. For the $\eta$ and $\eta^{\prime}$ only the VMD parametrization is considered.  \label{tab:fse2}}
\begin{center}
\begin{tabular}{lc|@{\quad}c@{\quad}c@{\quad}c@{\quad}c@{\quad}}
\hline
Id & $L~[\fm]$ & $v^{\pi;\mathrm{(LMD)}}$ & $v^{\pi;\mathrm{(VMD)}}$ & $v^\eta$ & $v^{\etap}$ \\
\hline
V3-L24 	& 	$3$ 	&	 $5.02(6)$ 	& 	3.69(4)	&  	$2.47(4)$ 		& $1.64(10)$  \\
V4-L32 	& 	$4$ 	&	 $1.30(6)$ 	& 	0.69(7)	& 	$0.60(5)$ 		& $0.65(10)$ \\
\hline
V3-L28 	& 	$3$ 	&	 $8.76(8)$ 	& 	6.56(4)	& 	$2.41(4)$ 		& $1.22(7)$ \\
V6-L56-1 	& 	$6$ 	&	 $0.43(26)$ 	& 	0.19(25)	& 	$0.16(20)$ 	& $0.30(16)$ \\
V6-L56-2 	& 	$6$ 	&	 $0.47(25)$ 	& 			& 	$''$ 	& $''$ \\
\hline
V3-L32 	& 	$3$ 	& 	$11.23(14)$ 	& 	8.83(6) 	& 	$2.43(4)$ 		& $1.51(4)$ \\
V6-L64 	& 	$6$ 	& 	$0.75(28)$ 	& 	0.48(32) 	& 	$0.11(18)$ 	& $0.15(27)$ \\
\hline
\end{tabular}
\end{center}
\end{table}

As a direct test of our procedure, we compare the results obtained for the connected contribution, using two different box sizes, before and after finite-volume correction. 
We emphasize that finite-size effects due to the pion-pole exchange are enhanced by a factor $34/9$ in the connected contribution compared to the total light quark contribution.
The results at three values of the lattice spacing are shown in \Table{tab:fse_check}. 
As expected, we observe that FSE increases as the lattice spacing decreases. 
At the finest lattice spacing, despite the large correction, the results in small and large volumes agree within $1.9~\sigma$. The deviation might be explained by the pion-loop contribution which is not considered here.  

Our final continuum extrapolation for the light-quark contribution relies solely on the large volume ensembles, where FSE corrections are small compared to the statistical precision. We observe that the LMD model provides a better description of the FSE, while the prediction using a VMD TFF is 20-25\% smaller. We thus associate a systematic uncertainty of 25\% on the FSE correction.

\begin{table}[b]
\renewcommand{\arraystretch}{1.25}
\caption{Test of our finite-volume correction estimate on the connected light-quark contribution. The spatial extent of the lattices are $L_1 \approx 3~\fm$ and  $L_2 \approx 6~\fm$, except at $a = 0.132~\fm$ where  $L_2 \approx 4~\fm$. The last column is the difference between the finite-volume corrected values in large and small volumes: $\Delta = (a_{\mu}^{\conn,l}(L_2) + \text{FSE}(L_2) ) - (a_{\mu}^{\conn,l}(L_1) + \text{FSE}(L_1))$. Numbers are given in units of $10^{-11}$.\label{tab:fse_check}}
\begin{center}
\begin{tabular}{l@{\quad}l@{\quad}c@{\quad}c@{\quad}c@{\quad}c}
\hline
$a~[\fm]$ 	& 	$a_{\mu}^{\conn,l}(L_1)$	&	FSE($L_1$)	&	$a_{\mu}^{\conn,l}(L_2)$	&	FSE($L_2$)	&	$\Delta$	\\
\hline
0.132 	& 	110.4(3.8)	&	19.0	&	117.6(4.6)		&	4.87		&	$-6.9(6.0)$ \\
\hline
0.112	&	118.8(2.9)	&	33.1	&	141.9(11.9)	&	1.74		&	$-8.3(12.2)$ 	\\
0.112	&			&	33.1	&	143.0(9.8)		&	1.74		&	$-7.2(10.2$) \\
\hline
0.095	&	117.0(3.8)	&	42.4	&	171.4(6.8)		&	2.83		&	$+14.8(7.8)$ \\
\hline
\end{tabular}
\end{center}
\end{table}

\subsubsection{Stategies }

\begin{figure}[t]
	\includegraphics*[width=0.49\linewidth]{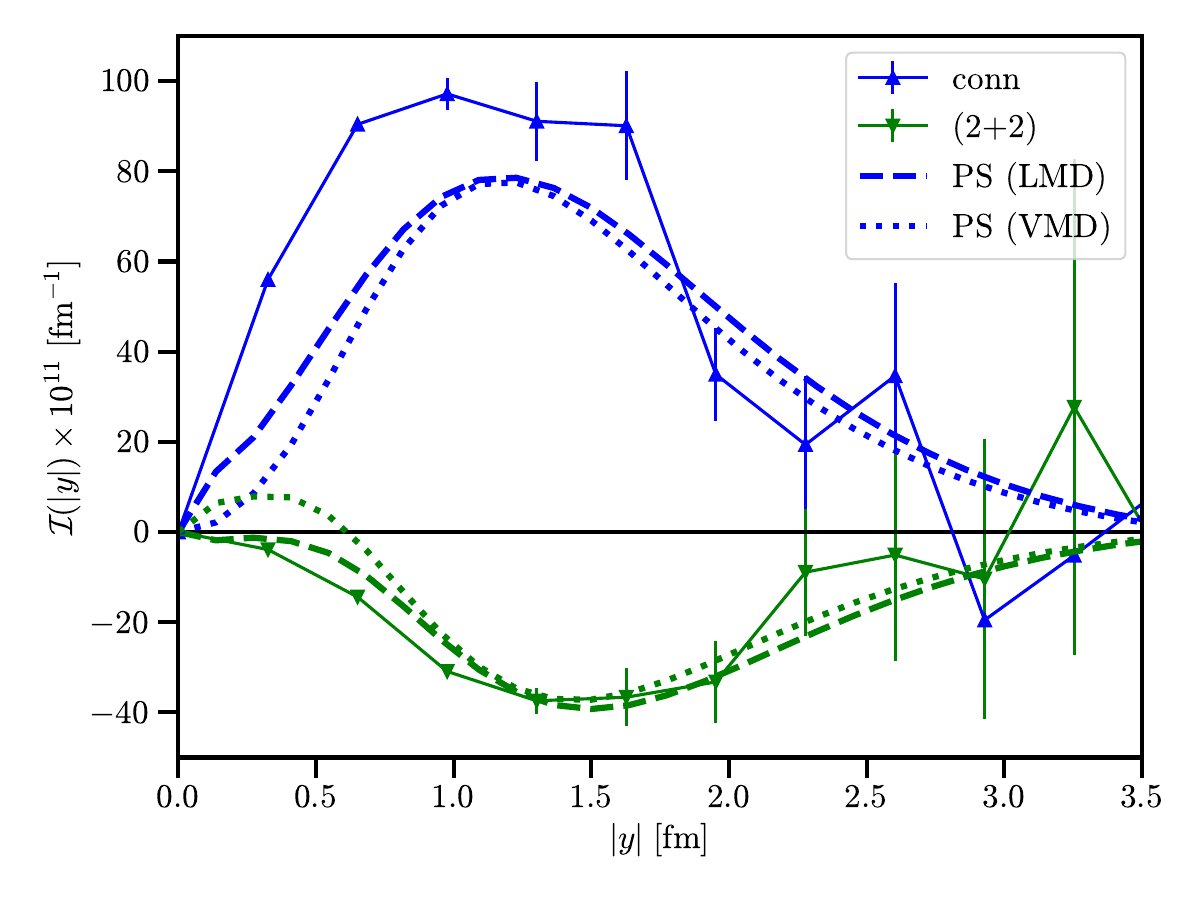}	
	\includegraphics*[width=0.49\linewidth]{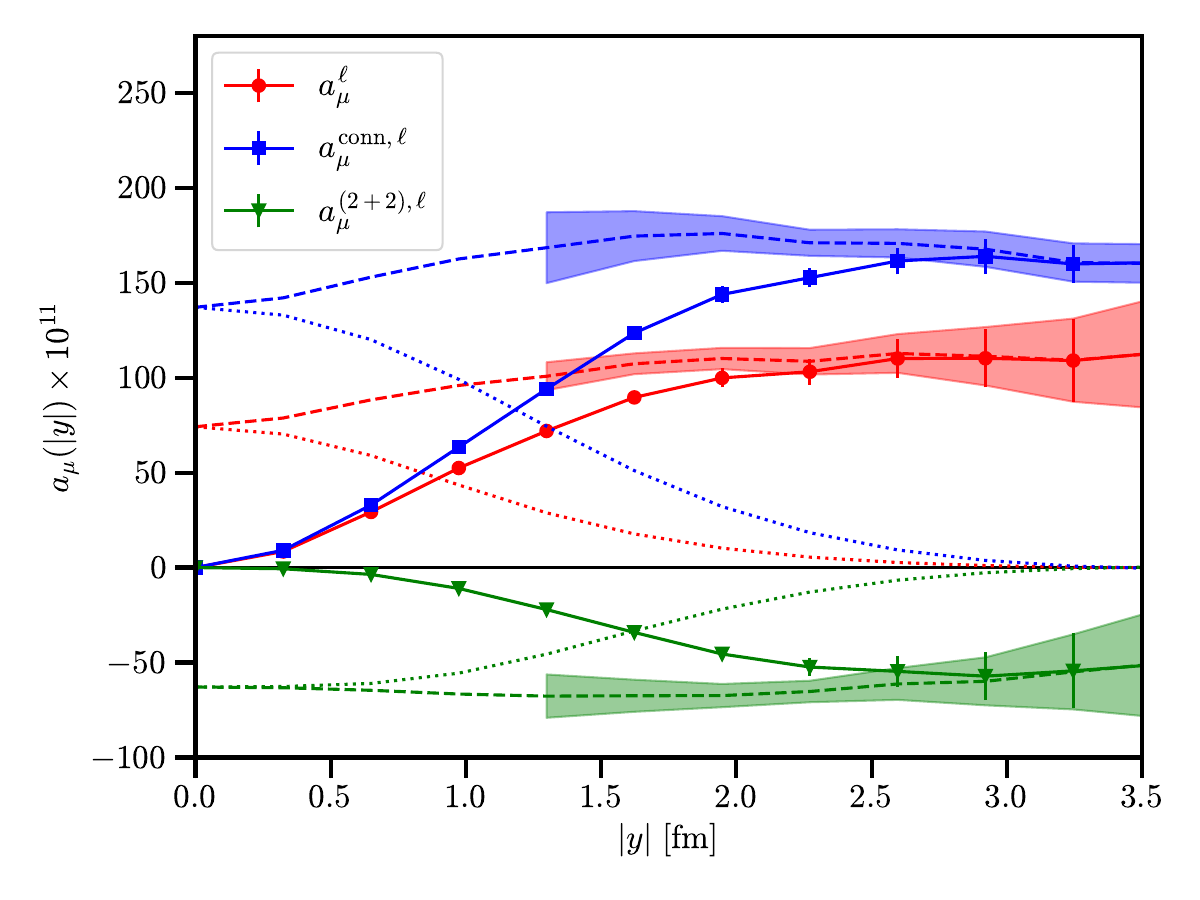}	
	\caption{Integrands (left) and partial sums (right) for the light quark contribution on the ensemble V6-L64 at our finest lattice spacing. On the left panel, the dashed lines represent the pseudoscalar-pole contribution in finite volume. On the right panel, the dotted line represents the partial sum for the pseudoscalar-pole contribution in the range $[|y|, \infty]$. The dashed line is the partial sum for lattice data up to $|y|$, supplemented by the pseudoscalar-pole contribution in the range $[|y|, \infty]$. The bands represent the sum of the statistical and systematic uncertainties.}
\label{fig:intlight}
\end{figure}

Our procedure to compute the light-quark contribution is the following. 
The bulk of the contribution with $|y| \leq |y|_{\rm cut}$ is obtained by integrating lattice data while the small tail of the integrand is estimated using the pseudoscalar-pole contribution in finite-volume presented in~\Section{eq:tail}. The finite-volume correction is then added for each ensemble, prior to the continuum extrapolation. 
Explicitly, for the quark-connected contribution, we compute 
\begin{align}
a^{\conn,l}_\mu = 
\int_0^{\ycut} \dd y\ \mathcal{I}^{\conn,l}(y) + \frac{34}{9} \int_{\ycut}^\infty \dd y\ \mathcal{I}^{\pi}(y)  + \frac{34}{9} v^{\pi} \,,
\end{align}
where the sum is performed using the trapezoidal rule. 
For the (2+2) quark-disconnected contribution, we follow the same strategy with the additional contributions from the heavier $\eta$ and $\etap$ mesons
\begin{align}
a^{(2+2),l}_\mu 
= 
\int_0^{\ycut} \dd y\ \mathcal{I}^{(2+2),l}(y) 
+ 
\int_{\ycut}^\infty \dd y\ 
\left[ 
\mathcal{I}^{\eta}(y) + \mathcal{I}^{\etap}(y) - \frac{25}{9} \mathcal{I}^{\pi}(y) 
\right]
+ v^{\eta} + v^{\etap} - \frac{25}{9} v^{\pi} \,.
\end{align}
Finally, for the total light-quark contribution, we evaluate
\begin{align}
a^{l}_\mu 
= 
\int_0^{\ycut} \dd y \left[ \mathcal{I}^{\conn,l}(y) + \mathcal{I}^{(2+2),l}(y) \right] + 
\int_{\ycut}^\infty \dd y 
\left[ \mathcal{I}^{\pi}(y) + \mathcal{I}^{\eta}(y) + \mathcal{I}^{\etap}(y) \right]
+ v^{\pi} + v^{\eta} + v^{\etap} \,.
\label{eq:light1}
\end{align}
In the right panel of~\Fig{fig:intlight}, we show our result for the partial sum as a function of $|y|$. To minimize any model dependence introduced by the tail reconstruction, $\ycut$ is chosen such that the tail contribution is comparable or smaller to the statistical error. 
The results for all ensembles, as well as the values of $\ycut$, are listed in Tables~\ref{tab:conn_light}, \ref{tab:disc_light} and~\ref{tab:tot_light}.

In~\cite{Blum:2023vlm}, it was noticed that adding $25/34$ of the connected contribution to the disconnected contribution should exactly cancel the pion-pole contribution at large distances where it is dominant. 
In that case, the integral converges much faster w.r.t.\ the upper integration limit. The missing contribution is then given by $a_\mu^{\conn}$ with the small factor $9/34$, and can be added later. Thus we evaluate 
 \begin{equation}
 a_\mu = a_\mu^{\rm no-pion} + \frac{9}{34} a_\mu^{\conn,l} + v^{\pi} + v^{\eta} + v^{\etap} 
 \label{eq:light2}
\end{equation}
with
\begin{align}
a_\mu^{\rm no-pion} = \int_0^{\ycut} \dd y \left[ \frac{25}{34} \mathcal{I}^{\conn,l}(y) + \mathcal{I}^{(2+2)}(y) \right] \,.
\end{align}
The results for $a_\mu^{\rm no-pion}$ are given in~\Table{tab:nopi}. \

\subsection{Continuum extrapolation}

\begin{figure}[t]
	\includegraphics*[width=0.49\linewidth]{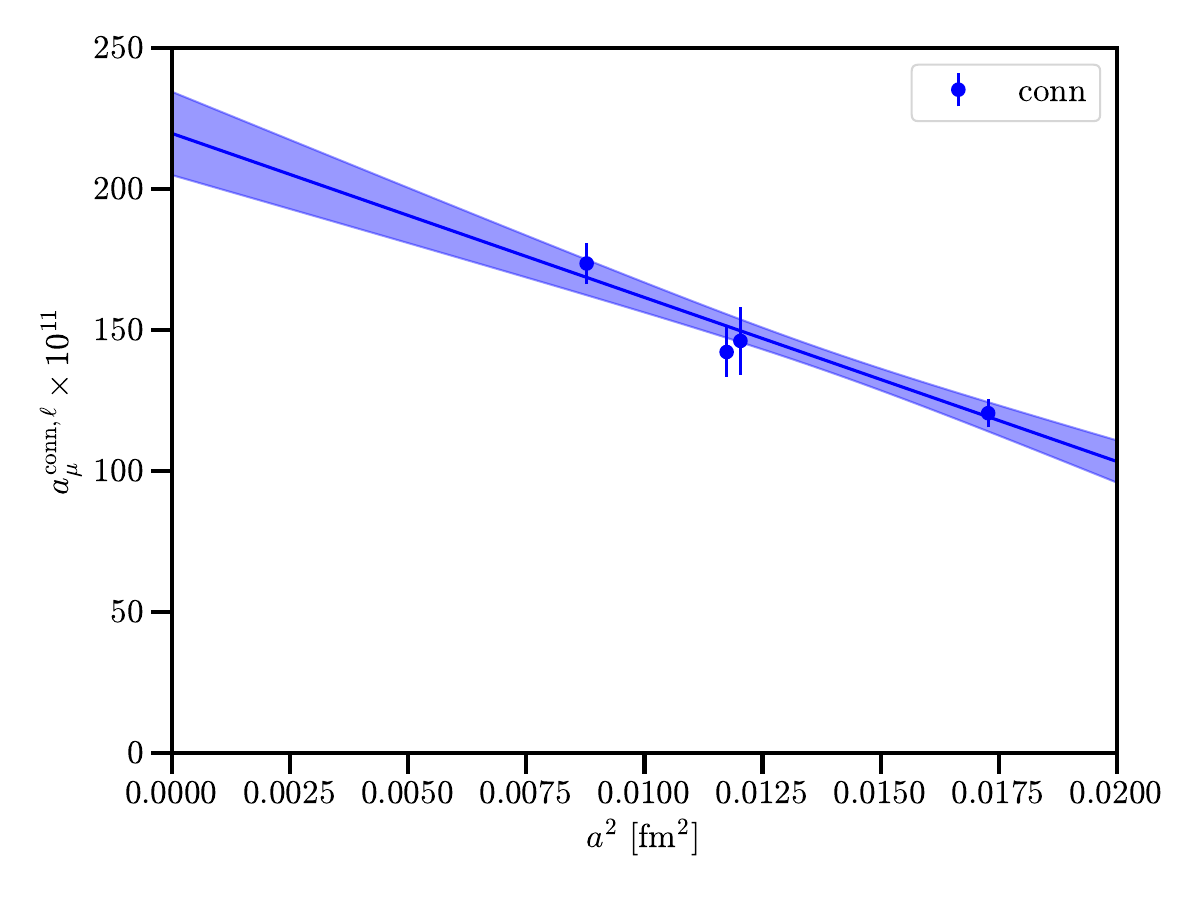}
	\includegraphics*[width=0.49\linewidth]{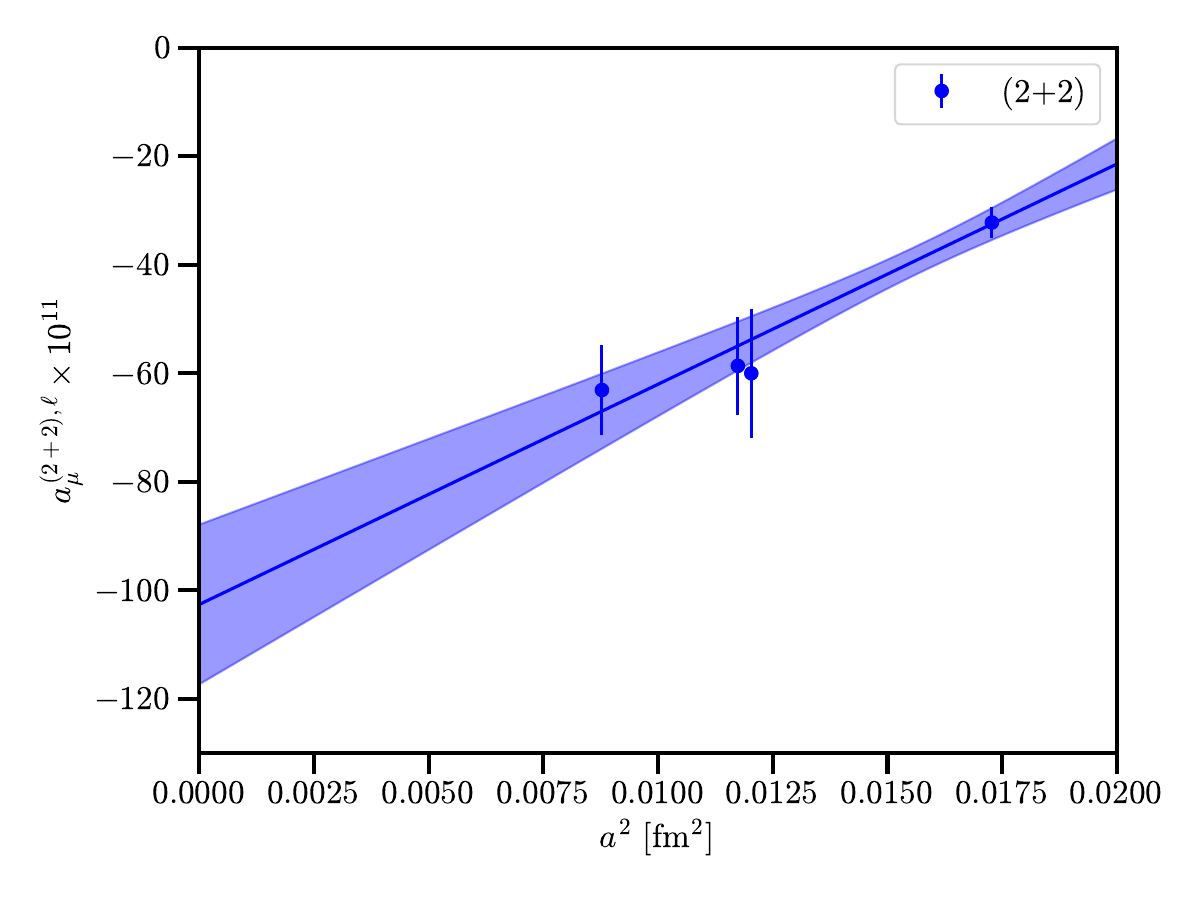}
	\includegraphics*[width=0.49\linewidth]{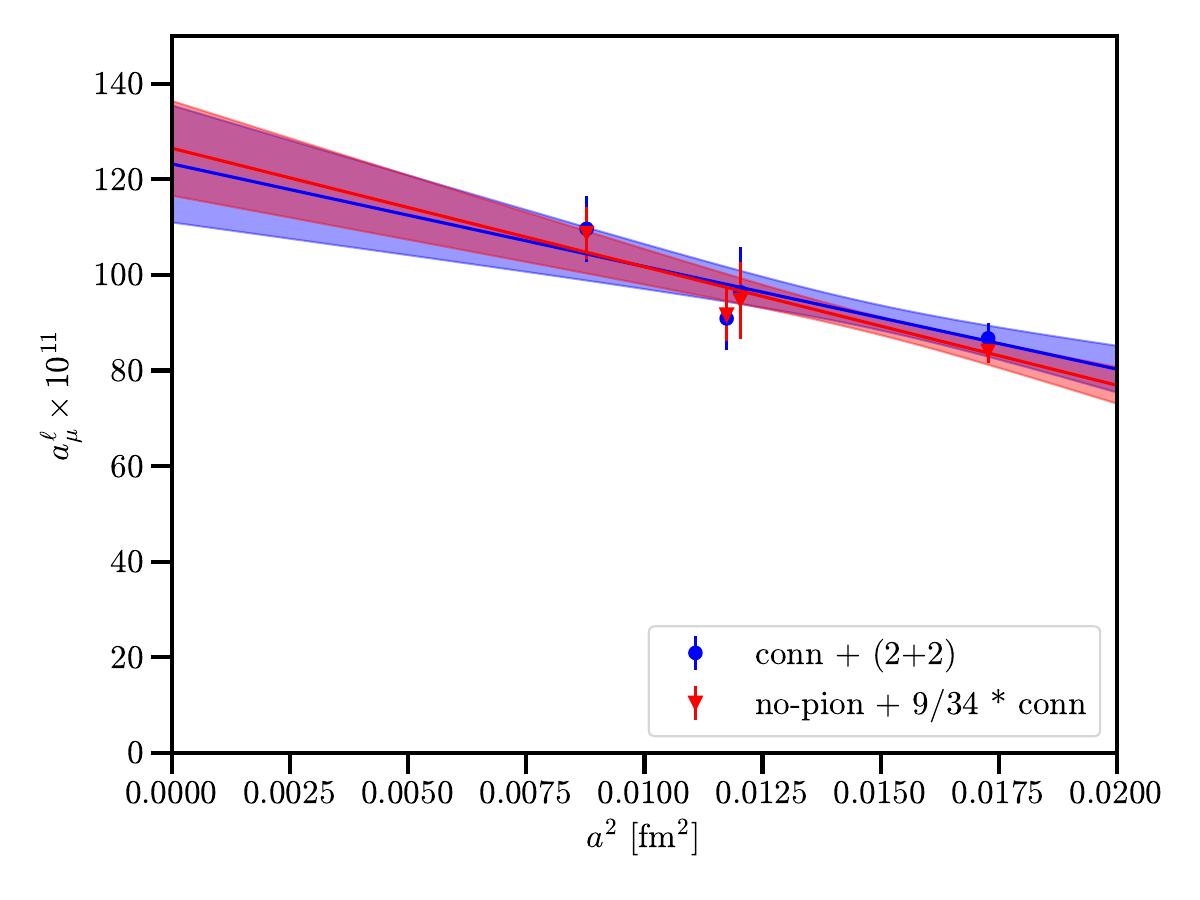}
	\caption{Top panel: Continuum extrapolations of the connected (left) and leading disconnected (right) light-quark contributions. Bottom panel: Continuum extrapolation of the total light-quark contribution using either~\Eq{eq:light1} (blue points) or \Eq{eq:light2} (red points).}
\label{fig:extrap_light}
\end{figure}

For each contribution, the continuum extrapolation is performed assuming discretization effects proportional to $a^2$. The results, which are shown in~\Fig{fig:extrap_light}, read
\begin{subequations}
\label{eq:light}
\begin{align}
a_{\mu}^{\conn,l} = 220.1(13.5) \times 10^{-11} \,,\\
a_{\mu}^{(2+2),l} = -101.1(12.8)  \times 10^{-11} \,,\\
a_{\mu}^{l} = 122.6(11.6) \times 10^{-11} \,,
\end{align}
\end{subequations}
where the total light-quark contribution is obtained using~\Eq{eq:light1} (see \Table{tab:tot_light}). Instead, using~\Eq{eq:light2} and the data summarized in~\Table{tab:nopi}, the continuum extrapolation for the light-quark contribution leads to the value $a_{\mu}^{l} = 126.2(9.2) \times 10^{-11}$. The correlated difference between both determinations is $3.6(2.7) \times 10^{-11}$. 

Our analysis is currently limited to three values of the lattice spacing. 
First, we note that for the strange and charm quark contributions, presented in the following sections, we also observe an excellent scaling in $a^2$ in the whole range of lattice spacings considered in this work. 
Second, to get further confidence that the extrapolation is under control, it is interesting to correct our data for the dominant source of taste-breaking effects: the pion-pole contribution. At our coarsest lattice spacing, subtracting the pion-pole contribution calculated using the TFF computed at the same values of $a$ and adding back the pion-pole contribution evaluated in the continuum limit in~\cite{Gerardin:2023naa}, the value is shifted from $a_{\mu}^{l} = 86.8 \times 10^{-11}$ to $a_{\mu}^{l} = 115.1\times 10^{-11}$, very close to our continuum extrapolated value in~\Eq{eq:light}.

\begin{table}
\renewcommand{\arraystretch}{1.23}
\caption{Connected light-quark contribution in units of $10^{-11}$. For all ensembles considered, the last column is the sum of the $|y| \leq \ycut$, $|y| > \ycut$ contributions, including the finite-size effect (FSE) correction.
\label{tab:conn_light}}
\begin{center}
\begin{tabular}{l@{\quad}l@{\quad}c@{\quad}c@{\quad}c@{\quad}c}
\hline
Id 		& 	$\ycut$	&	$a_{\mu}^{\conn}({|y| \leq \ycut})$	&	$a_{\mu}^{\conn}({|y| > \ycut})$	&	FSE	&	$a_{\mu}^{\conn,l}$		\\
\hline
V4-L32 	& 	2.63 fm	&	$115.5(4.6)_{\stat}$		&	2.1(0.1)	&	4.9(1.0)	&	$122.5(4.6)_{\stat}(1.0)_{\syst}$ \\ 
\hline
V6-L56-1	&	2.66 fm 	&	$136.8(11.8)_{\stat}$		&	5.1(1.1)	&	1.7(0.3)	&	$\ 143.6(11.8)_{\stat}(1.1)_{\syst}$\\ 
V6-L56-2	&	2.63 fm 	&	$137.6(9.8)_{\stat}$		&	5.4(1.1)	&	1.7(0.3)	&	$144.7(9.8)_{\stat}(1.1)_{\syst}$ \\
\hline
V6-L64	&	2.60 fm 	&	$162.0(6.8)_{\stat}$		&	9.4(1.7)	&	2.8(0.6)	&	$174.2(6.8)_{\stat}(1.8)_{\syst}$ \\ 
\hline
\end{tabular}
\end{center}
\caption{Same as~\Table{tab:conn_light} for the leading disconnected light-quark contribution in units of $10^{-11}$.  
\label{tab:disc_light}}
\begin{center}
\begin{tabular}{l@{\quad}l@{\quad}c@{\quad}c@{\quad}c@{\quad}c}
\hline
Id 	& 	$\ycut$		&	$a_{\mu}^{(2+2)}({|y| \leq \ycut})$	&	$a_{\mu}^{(2+2)}({|y| > \ycut})$	&	FSE	&	$a_{\mu}^{(2+2),l}$		\\
\hline
V4-L32 	& 	2.63 fm	&	$-28.1(2.7)_{\stat}$		&	$-1.5(0.1)$	&	$-2.3(0.5)$	&	$-32.0(2.7)_{\stat}(1.1)_{\syst}$ \\
\hline
V6-L56-1	&	2.28 fm 	&	$-44.6(7.3)_{\stat}$		&	$-8.6(2.2)$	&	$-0.8(0.2)$	&	$\ -53.9(7.3)_{\stat}(2.2)_{\syst}$ \\ 
V6-L56-2	&	2.63 fm 	&	$-53.7(8.2)_{\stat}$		&	$-3.8(0.8)$	&	$-0.8(0.2)$	&	$-58.3(8.2)_{\stat}(0.8)_{\syst}$ \\
\hline
V6-L64	&	2.60 fm 	&	$-54.3(8.1)_{\stat}$		&	$-6.7(1.2)$	&	$-1.8(0.4)$	&	$-62.9(8.1)_{\stat}(1.3)_{\syst}$ \\
\hline
\end{tabular}
\end{center}
\caption{Same as~\Table{tab:conn_light} for the total light-quark contribution in units of $10^{-11}$.  
\label{tab:tot_light}}
\begin{center}
\begin{tabular}{l@{\quad}l@{\quad}c@{\quad}c@{\quad}c@{\quad}c}
\hline
Id 	& 	$y_{\rm cut}$	&	$a_{\mu}^{\mathrm{tot}}({|y| \leq \ycut})$	&	$a_{\mu}^{\mathrm{tot}}({|y| > \ycut})$	&	FSE	&	$a_{\mu}^{\mathrm{tot},l}$		\\
\hline
V4-L32 	& 	2.37 fm	&	$82.9(3.2)	_{\stat}$		&	1.4(0.1)	&	2.5(0.5)	&	$86.8(3.2)_{\stat}(0.5)_{\syst}$ \\  
\hline
V6-L56-1	&	2.28 fm	&	$91.7(9.6)_{\stat}$		&	3.9(0.6)	&	1.0(0.2)	&	$96.6(9.6)_{\stat}(0.6)_{\syst}$ \\ 
V6-L56-2	&	2.25	 fm 	&	$86.1(6.6)_{\stat}$		&	4.1(0.6)	&	1.0(0.2)	&	$91.2(6.6)_{\stat}(0.6)_{\syst}$\\
\hline
V6-L64	&	2.27 fm 	&	$102.4(6.7)_{\stat}$		&	5.5(0.8)	&	1.0(0.2)	&	$108.9(6.7)_{\stat}(0.8)_{\syst}$ \\
\hline
\end{tabular}
\end{center}
\caption{Value of $a^{\rm no-pion}_{\mu}$ in units of $10^{-11}$ for all ensembles considered.  
\label{tab:nopi}}
\begin{center}
\begin{tabular}{l@{\quad}l@{\quad}c}
\hline
Id 	& 	$y_{\rm cut}$		&	$a^{\rm no-pion}_{\mu}$ \\
\hline
V4-L32 	& 	2.10 fm		&	$49.3(2.0)_{\stat}$ \\
\hline
V6-L56-1	&	1.90 fm 		&	$55.6(4.7)_{\stat}$ \\
V6-L56-2	&	1.88 fm		&	$53.2(4.4)_{\stat}$ \\  
\hline
V6-L64	&	1.95 fm 		&	$59.3(3.7)_{\stat}$ \\
\hline
\end{tabular}
\end{center}
\end{table}

\section{Strange-quark contribution \label{sec:strange}}

\subsection{Connected contribution}

\begin{figure}[t]
	\includegraphics*[width=0.47\linewidth]{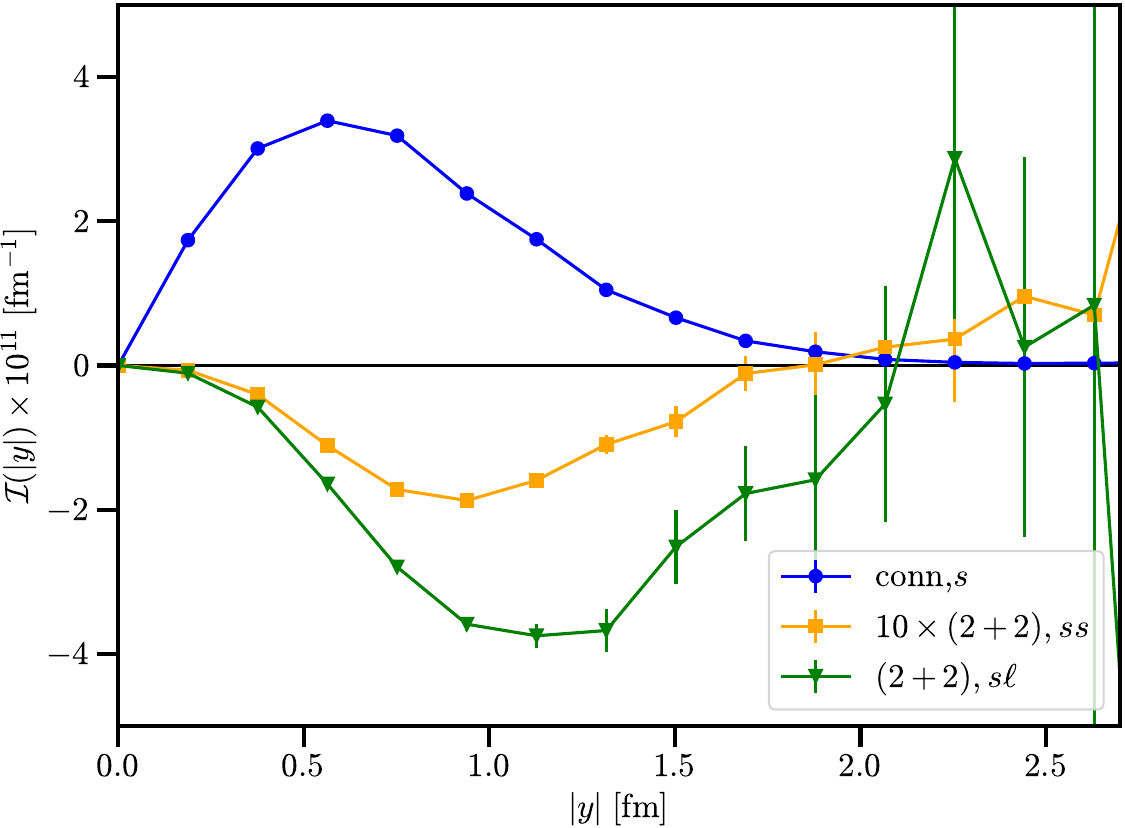} \quad
	\includegraphics*[width=0.47\linewidth]{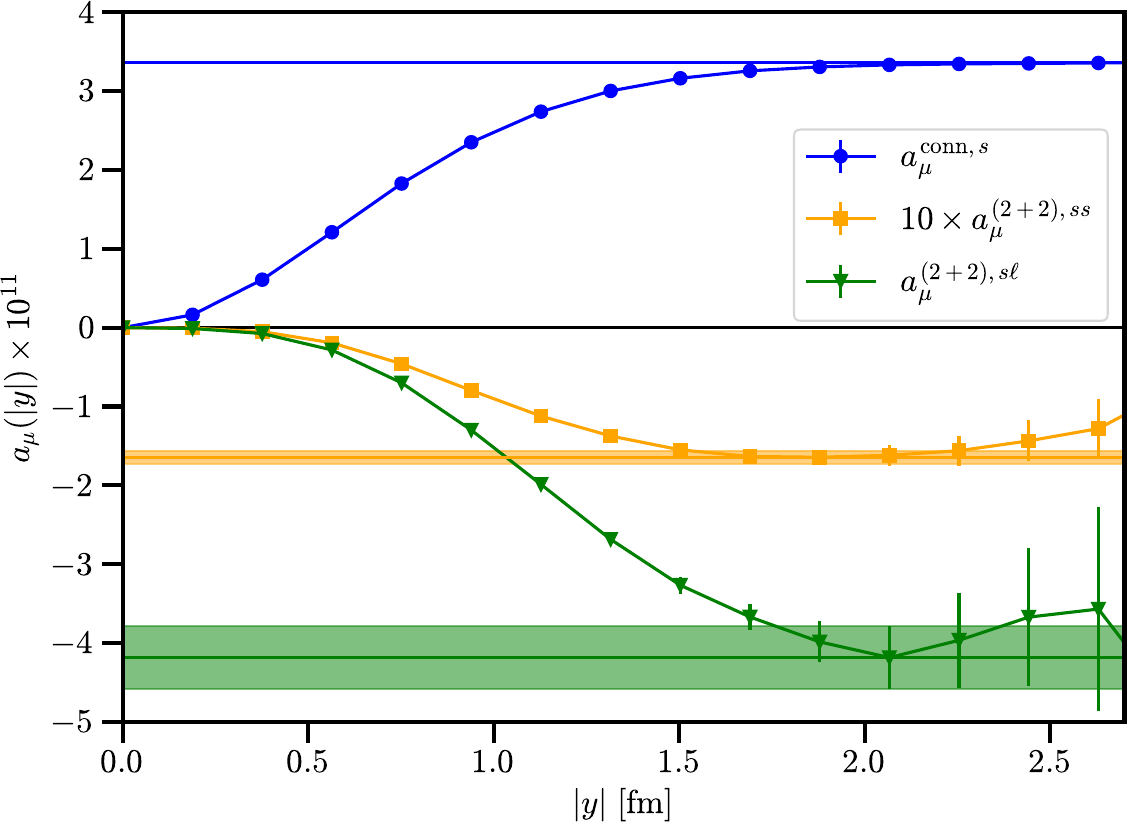}
	\caption{Left: integrands for the strange-quark contributions at $a = 0.095$~fm. Right: partial sums for the strange-quark contributions at the same lattice spacing. The disconnected strange-strange contribution has been multiplied by 10 for visibility. The horizontal lines and bands represent the value of the corresponding fully-summed contributions and its statistical uncertainty.}
\label{fig:amuS}
\end{figure}

The strange-quark contribution includes all diagrams with at least one valence strange-quark but no valence charm-quark. The latter are included in the charm-quark contribution discussed in \Section{sec:charm}. 
Compared to the light-quark contribution, the integrand is less long-range and does not suffer from the bad signal-to-noise ratio (see~\Fig{fig:amuS}). No cut in the $x$-integration or modeling of the tail is needed and the results for individual ensembles are summarized in \Table{tab:dataS}. A statistical precision below 1\% is reached on all ensembles. 

However, systematic effects that were negligible for the light-quark contribution become relevant here. First, we need to take into account the fact that our simulations are not performed exactly at the physical pion and kaon masses. Thus, at several values of the lattice spacings, we have generated data using slightly different values of the pion and kaon masses, leading to a total of 11 ensembles. 
No dependence on the pion mass mistuning is observed and we use the ratio
$\delta M_{ss} / M_{ss}^{\phys}$
as a proxy for the slight mistuning of the strange-quark masses. Here, $M_{ss}$ is the mass from the connected pseudoscalar two-point correlation function and we use the physical value $M_{ss}^{\phys} = 689.89(49)~\MeV$ computed at the physical point in~\cite{Borsanyi:2020mff}.

Second, we observe that the uncertainty associated with the lattice spacing is not negligible. The scale enters through the muon mass both explicitly in the prefactor and in the QED kernel function of \Eq{eq:master}. Since the contraction of the QED weight function with the four-point function is performed during the simulation, it is not trivial to propagate the scale uncertainty. 
On one ensemble, at our coarsest lattice spacing, we have performed two runs that differ only by the value of the lattice spacing (it is increased by 1\% in the simulation). This gives us access to the parametric derivative with respect to the lattice spacing
\begin{align}
\frac{\delta a_\mu(a)}{a_\mu(a)} = \frac{1}{a_\mu(a)} \frac{\partial a_\mu(a)}{\partial a} \delta a \approx \frac{1}{a_\mu(a)} \frac{\Delta a_\mu(a)}{\Delta a} \delta a = 1.6 \, \delta a \,.
\end{align}
This scale uncertainty is added quadratically to the statistical uncertainty. 

\begin{table}[b]
\renewcommand{\arraystretch}{1.25}
\caption{Connected and leading disconnected strange-quark contribution for each individual ensemble. The first error is statistical, the second error is due to the scale uncertainty. The disconnected contribution is computed only on a subset of the ensembles. \label{tab:dataS}}
\begin{center}
\begin{tabular}{l@{\quad}c@{\quad}c@{\quad}c}
\hline
Id 	& 	$a^{\conn,s}_\mu \times 10^{11}$		& 	$a^{(2+2),sl}_\mu \times 10^{11}$		& 	$a^{(2+2),ss}_\mu \times 10^{11}$\\
\hline
V3-L24 	& 	$3.061(2)_{\stat}(11)_a$	&	$-2.88(17)_{\stat}$	&	$-0.133(6)_{\stat}$	\\
V4-L32 	& 	$3.087(2)_{\stat}(11)_a$	&	$-3.21(29)_{\stat}$	&	$-0.134(15)_{\stat}$	\\
V6-L48 	& 	$3.105(17)_{\stat}(11)_a$	&	-	&	-		\\
\hline
V3-L28 	& 	$3.264(2)_{\stat}(15)_a$	& 	$-3.40(33)_{\stat}$	&	$-0.151(7)_{\stat}$	\\
V6-L56-1	&	$3.270(14)_{\stat}(6)_a$	&	-	&	-		\\
\hline
V3-L32-1 	& 	$3.339(7)_{\stat}(12)_a$	&	-	&	-		\\
V3-L32-2 	& 	$3.363(3)_{\stat}(16)_a$	& 	$-4.18(40)_{\stat}$	&	$-0.164(8)_{\stat}$	\\
V3-L32-3 	& 	$3.310(8)_{\stat}(15)_a$	&	-	&	-		\\
\hline
V3-L40-1 	& 	$3.473(5)_{\stat}(3)_a$	&	$-4.15(61)_{\stat}$	&	$-0.200(23)_{\stat}$	\\
\hline
V3-L48 	& 	$3.538(9)_{\stat}(5)_a$	&	-	&	-		\\
V3-L48 	& 	$3.513(9)_{\stat}(4)_a$	&	-	&	-		\\
\hline
\end{tabular}
\end{center}
\end{table}

\begin{figure}[t]
	\includegraphics*[width=0.47\linewidth]{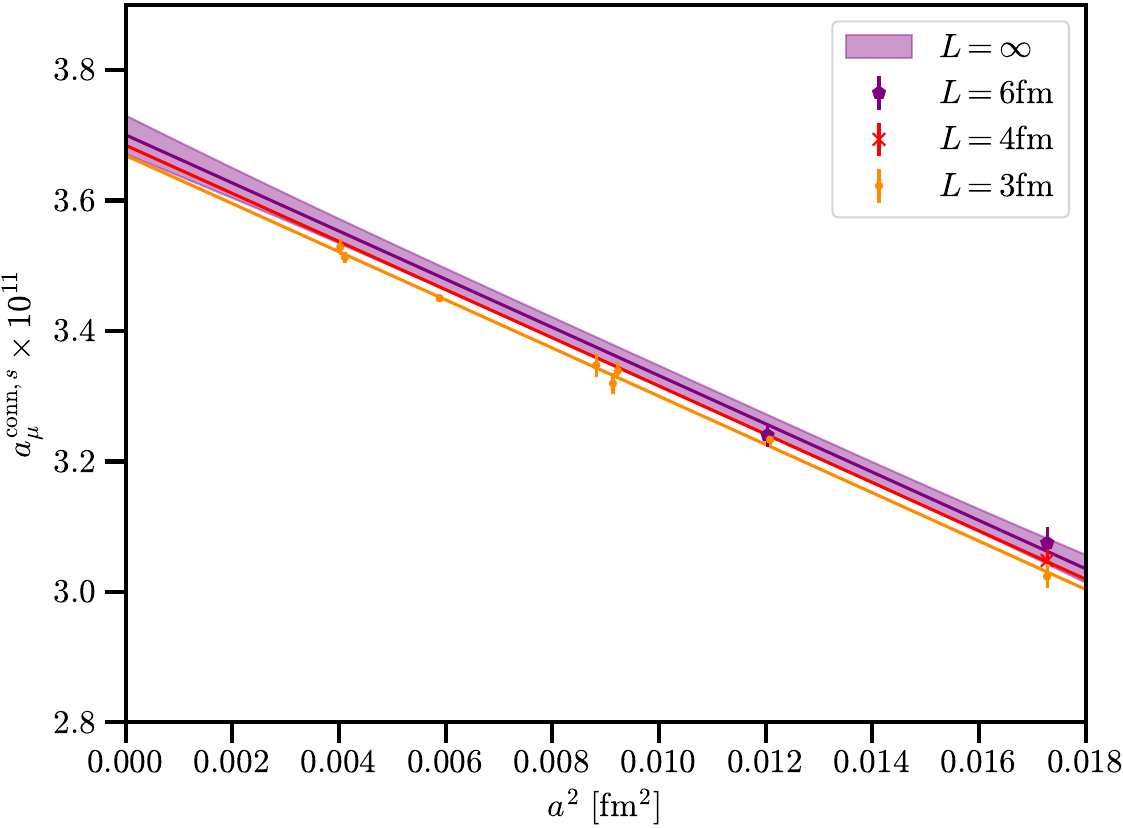} \quad
	\includegraphics*[width=0.47\linewidth]{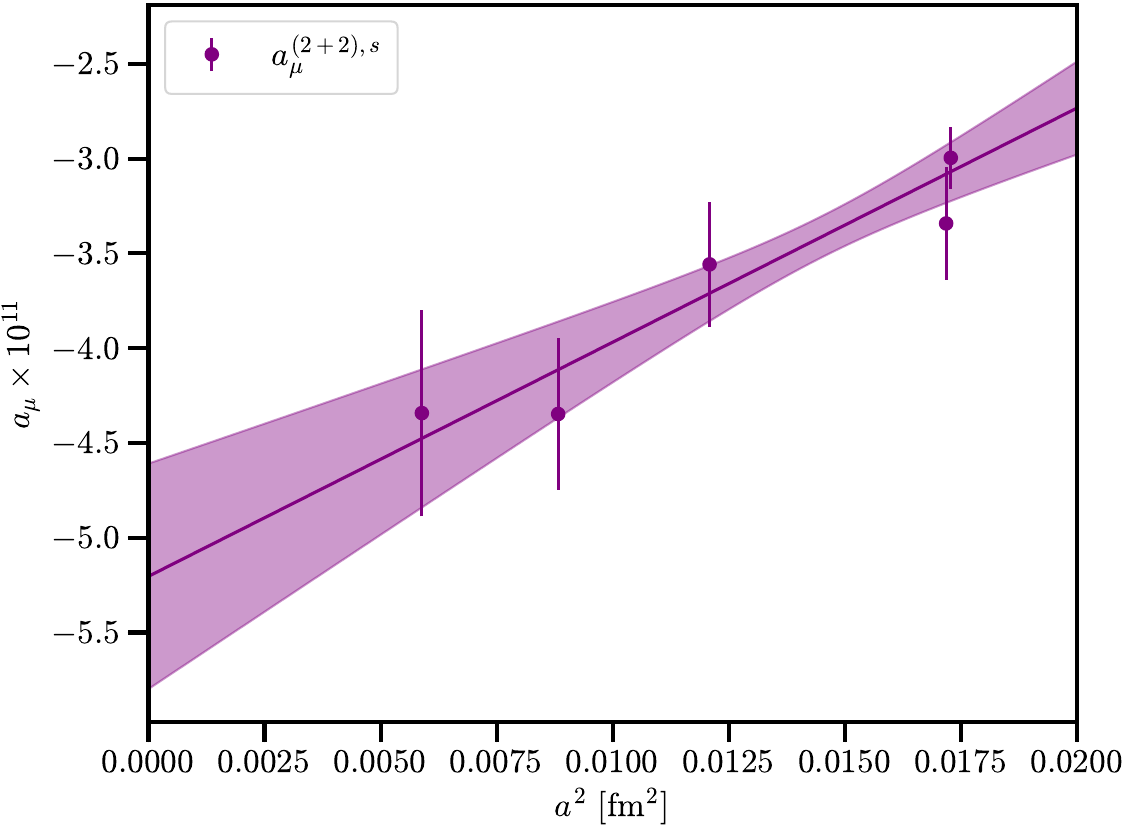}
	\vspace{-0.2cm}
	\caption{Continuum extrapolation for the connected strange-quark contribution (left) and for the leading strange-quark $(2+2)$ disconnected contribution (right).}
\label{fig:extrapS}
\end{figure}

To better control the continuum extrapolation, we have generated data at five values of the lattice spacing using smaller volumes with $L \approx 3~$fm. Although finite-size effects are small, they are statistically significant and additional large volume ensembles are used to correct for these effects. 
The continuum and infinite-volume extrapolations are performed simultaneously assuming the functional dependence 
\begin{equation}
a_\mu(a^2, M_{ss}) = a_{\mu}^{\conn,s} + \beta_2 (\Lambda a)^2 + \delta_2\ (\Lambda a)^2\ \alpha_s^n(a^{-1}) + \beta_4 (\Lambda a)^4  + \gamma \frac{\delta M_{ss}}{M_{ss}^\phys} + \lambda\ e^{- L m_{\pi}} \,,
\label{eq:fitS}
\end{equation}
where $n=2$ or 3 and $\Lambda = 0.5~$GeV is a typical QCD scale. Here,  $\alpha_s(a^{-1})$ is the strong coupling constant in the $\overline{\rm MS}$ scheme evaluated at the scale $1/a$ and $m_{\pi} = 135~$MeV is the physical pion mass.  
In practice, we have performed several continuum extrapolations that differ by the set of fit parameters that are allowed to vary. In the most agressive case, which corresponds to $\delta_2=0$ and $\beta_4=0$, we also perform a cut in the lattice spacing. The list of models and the corresponding extrapolations are given in \Table{tab:fitS}. The final result, obtained by a flat average over all models, reads
\begin{equation}
a_{\mu}^{\conn,s} = 3.694(25)_{\stat}(8)_{\syst} \times 10^{-11} \,,
\end{equation}
where the systematic uncertainty is estimated by computing the root mean square deviation of the fit results compared to the flat average. 
 A typical continuum extrapolation is shown in the left panel of~\Fig{fig:extrapS}.

\begin{table}[b]
\renewcommand{\arraystretch}{1.25}
\caption{Continuum extrapolations of the strange-quark contributions. A ``-'' means that the fit parameter is not included. Ensembles with lattice spacings larger than the cut are excluded from the fit. The last line is our final estimate using a model averaging with a flat weight over all variations. The systematic uncertainty is estimated by computing the root mean square deviation of the fit results compared to the average. \label{tab:fitS}}
\begin{center}
\begin{tabular}{c@{\quad}c@{\quad}c@{\quad}|@{\quad}c@{\quad}c@{\quad}|@{\quad}c@{\quad}c}
\hline
$\beta_4$ & $\delta_2$ & 	Cut & $a_{\mu}^{\conn,s} \times 10^{11}$ & $\chi^2/\dof$ & $a_{\mu}^{\rm (2+2),s} \times 10^{11}$ & $\chi^2/\dof$  \\
\hline
  -	& - 		&  -			&	$3.700(21)$	&	0.49	&	$-5.2(0.6)$ 	& 	0.51		\\ 
  -	& - 		&  0.12~fm	&	$3.678(30)$	& 	0.52 	&	$-5.5(1.1)$	&	0.49	 	\\
 yes  & - 		&  -			& 	$3.698(27)$	& 	0.57 &	 \\
  -  	& $n=2$	&  -			& 	$3.698(28)$	& 	0.57 	&				&			\\
  -  	& $n=3$	&  -			& 	$3.698(27)$	& 	0.57	&				& 		 	\\
\hline
\multicolumn{3}{c@{\quad}|@{\quad}}{ Model averaging} &	\multicolumn{2}{c@{\quad}|@{\quad}}{  $3.694(25)_{\stat}(8)_{\syst}$  }		&	\multicolumn{2}{c}{ $-5.4(0.8)_{\stat}(0.2)_{\syst}$ } 	\\
\hline
\end{tabular}
\end{center}
\end{table}

\subsection{Leading quark-disconnected contribution}

In \Fig{fig:amuS}, we observe that the pure strange-strange (2+2) disconnected contribution is about 20 times smaller than the strange-light (2+2) disconnected contribution. The strange-light contribution is negative and largely cancels the connected contribution such that the overall strange-quark contribution turns out to be negative. 
A clear signal is observed and we use respectively $y_{\rm cut} \approx 2.1~\fm$ and $y_{\rm cut} \approx 1.9~\fm$ for the light-strange and pure strange disconnected contribution when evaluating~\Eq{eq:amuy}. The data is statistically less precise and finite-size effects are neglected. 

This contribution has been computed only on a subset of our ensembles and the results are listed in \Table{tab:dataS}. 
The effect of the mistuning of the strange-quark, which is expected to be small compared to the statistical uncertainty, is neglected. Thus, the continuum extrapolation is performed using~\Eq{eq:fitS} setting $\gamma=\beta_4=\delta_2=\lambda=0$ and two different cuts in the lattice spacing. Our final estimate reads
\begin{equation}
a_{\mu}^{\rm (2+2),s} = -5.4(0.8)_\stat(0.2)_\syst \times 10^{-11} \,.
\end{equation}

\section{Charm-quark contribution \label{sec:charm}} 

\subsection{Connected contribution}

\begin{figure}[t]
	\includegraphics*[width=0.47\linewidth]{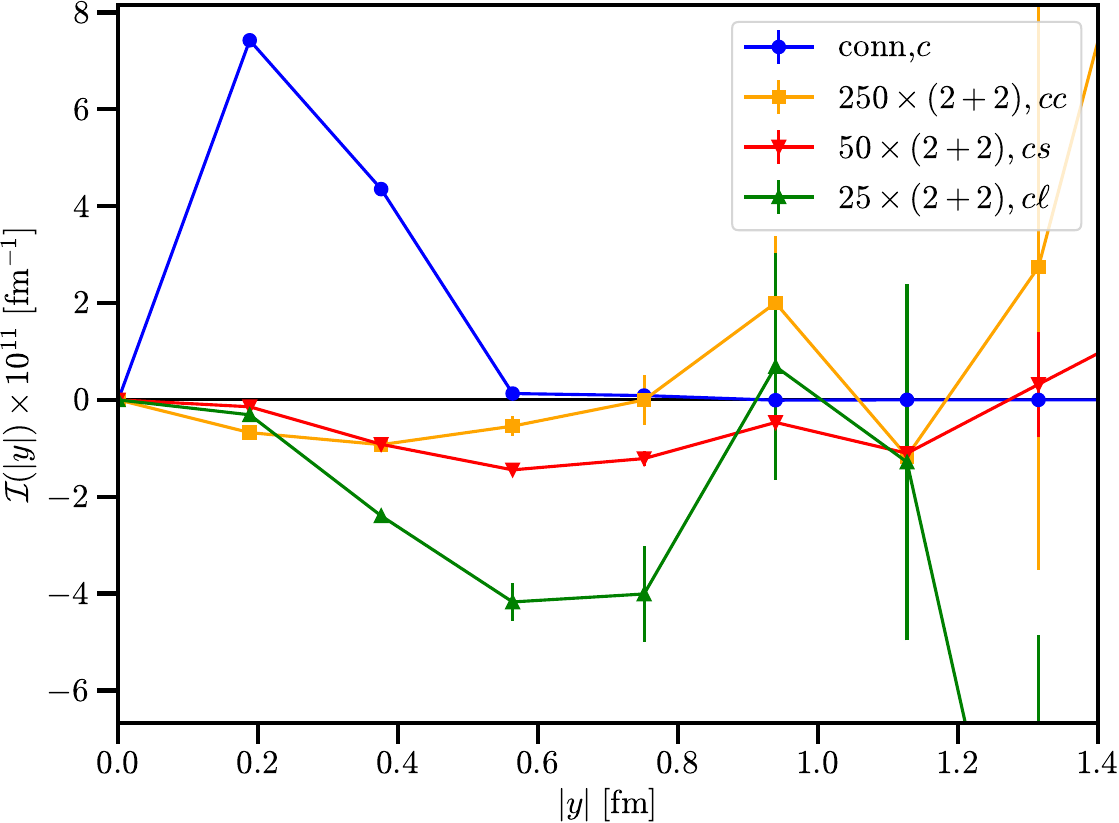} \quad
	\includegraphics*[width=0.47\linewidth]{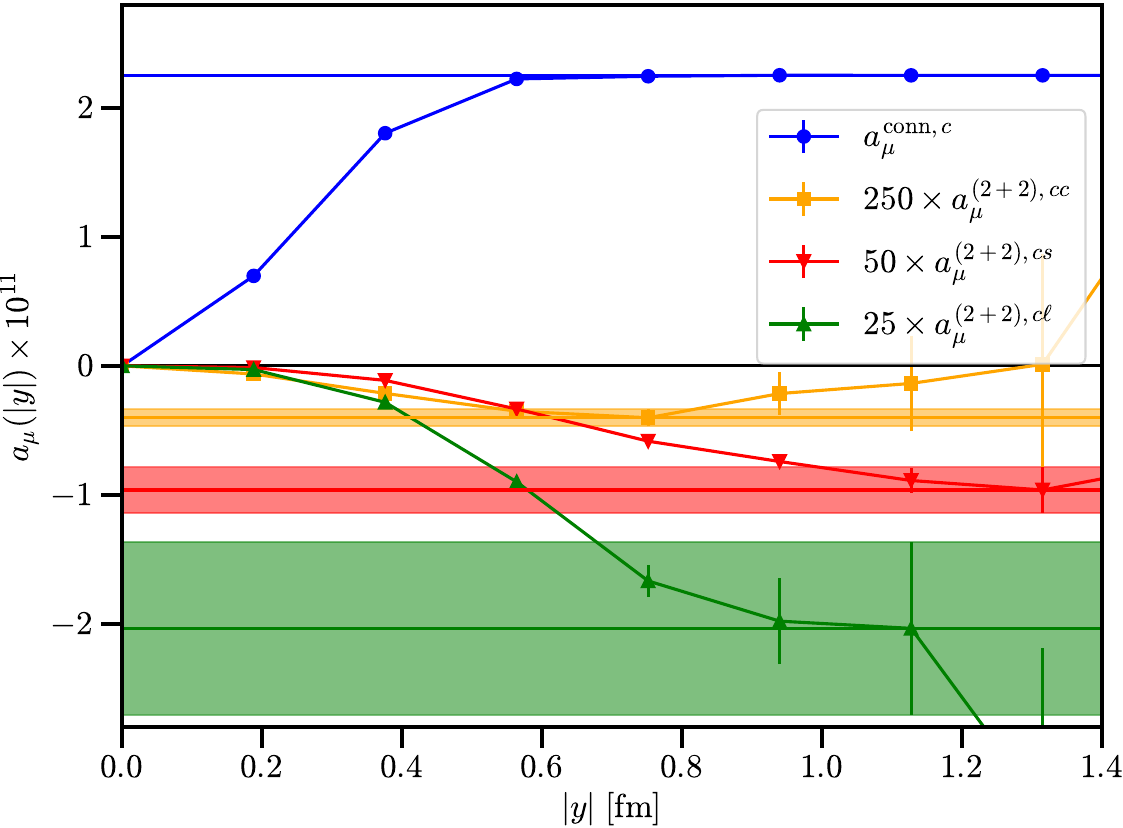}
	\caption{Left: Integrands for the charm-quark contribution at $a = 0.095$~fm. Right: Partial sums for the charm-quark contributions at the same lattice spacing. The disconnected charm-charm, charm-strange and charm-light contributions have been multiplied by 250, 50 and 25 respectively for visibility. The horizontal lines and bands represent the value of the corresponding fully-summed contributions and its statistical uncertainty.}
\label{fig:integrandC}
\end{figure}

\renewcommand{\arraystretch}{1.25}
\begin{table}[b]
\caption{Connected and leading disconnected charm-quark contributions for all ensembles and for both trajectories $\vec{n}$ used to sample the integrand. The first error is statistical, the second error is due to the scale uncertainty. \label{tab:dataC}} 
\begin{center}
\begin{tabular}{l@{\quad}c@{\quad}c@{\quad}c@{\quad}c@{\quad}c}
\hline
 & \multicolumn{2}{@{\quad}c@{\quad}}{ $a_{\mu}^{\conn;c} \times 10^{11}$ } & $a^{(2+2),cl}_\mu \times 10^{11}$ & $a^{(2+2),cs}_\mu \times 10^{11}$ & $a^{(2+2),cc}_\mu \times 10^{11}$ \\ 
Id		&	$\vec{n} = (1,1,1,1)$		&	$\vec{n} = (0,1,1,1)$		&	\multicolumn{3}{@{\quad}c@{\quad}}{ $\vec{n} = (1,1,1,1)$ } \\ 
\hline
V3-L24	&  $3.6102(5)_\stat(80)_{a}$	&  $3.5505(64)_\stat(79)_{a}$	&  $-0.055(20)_{\stat}$ 	&  $-0.0081(13)_{\stat}$ 	&  $-0.00100(11)_{\stat}$ \\
V4-L32	&  $3.6103(7)_\stat(80)_{a}$	&						&  $-0.050(31)_{\stat}$ 	&  $-0.0068(18)_{\stat}$  	&  $-0.00111(20)_{\stat}$ \\
\hline
V3-L28	&  $3.6988(8)_\stat(105)_{a}$	&  $3.6213(49)_\stat(103)_{a}$	&  $-0.110(36)_{\stat}$ 	&  $-0.0132(40)_{\stat}$ 	&  $-0.00095(34)_{\stat}$ \\
\hline
V3-L32-1	&  $3.5272(37)_\stat(80)_{a}$	&  $3.4660(47)_\stat(79)_{a}$	& 	-				&	-	&	-  \\
V3-L32-2	&  $3.4964(40)_\stat(97)_{a}$	&  $3.4387(52)_\stat(95)_{a}$	&  $-0.081(27)_{\stat}$	&  $-0.0192(35)_{\stat}$ 	&  $-0.00161(27)_{\stat}$ \\
V3-L32-3	&  $3.6951(9)_\stat(109)_{a}$	&  $3.6100(54)_\stat(107)_{a}$	& 	-				&	-	&	- \\
V3-L32-4	&  $3.8607(38)_\stat(108)_{a}$	&  $3.7543(62)_\stat(105)_{a}$ 	& 	-				&	-	&	- \\
\hline
V3-L40-1	&  $3.8367(16)_\stat(23)_{a}$	&  $3.7614(50)_\stat(23)_{a}$	&  $-0.136(32)_{\stat}$	&  $-0.0216(47)_{\stat}$   &. $-0.00355(89)_{\stat}$ \\
V3-L40-2	&  $3.7073(35)_\stat(18)_{a}$	&  $3.6311(47)_\stat(18)_{a}$	& 	-				&	-				&	- \\
\hline
V3-L48-1	&  $3.7518(29)_\stat(33)_{a}$	&  $3.6806(49)_\stat(32)_{a}$	& 	-				&	-				&	- \\
V3-L48-2	&  $3.6537(31)_\stat(24)_{a}$	&  $3.5912(41)_\stat(24)_{a}$	& 	-				&	-				&	- \\
\hline
V6-L128	&  $3.807(39)_\stat(4)_a$		&	-					&	-				&	-				&	- \\
\hline 
\end{tabular}
\end{center}
\end{table}

The connected charm-quark contribution is statistically very precise and finite-size effects are found to be negligible. As for the connected strange-quark contribution, no modeling of the tail is needed. The difficulty lies in the continuum extrapolation. 
Thus, in addition to our standard trajectory $\vec{n} = (1, 1, 1, 1)$, we add a second trajectory with $\vec{n} = (0, 1, 1, 1)$ to sample the integrand as a function of $|y|$. This trajectory has a smaller step size $\dd y = \sqrt{3}a$ which allows for a better sampling of the integrand\footnote{Trajectories with even smaller step sizes (using $\vec{n} = (0, 0, 1, 1)$ or even $\vec{n} = (0, 0, 0, 1)$) have been tested but are affected by very large discretization effects.}. Both trajectories are expected to agree in the  continuum limit. 
Second, we use tree-level improvement (TLI) to correct our data. We apply the correction
\begin{align}
a_\mu^{\mathrm{impr}}(a) = a_\mu^{\mathrm{latt}}(a) - a_\mu^{\mathrm{free}}(a) + a_\mu^{\mathrm{cont,free}}
\end{align}
where $a_\mu^{\mathrm{cont,free}} = 3.08592 \times 10^{-11}$ is the quark-loop contribution in the continuum limit at the physical charm quark mass (using the \MS\ value $\overline{m}_c(\overline{m}_c) = 1.27~\GeV$ from~\cite{ParticleDataGroup:2022pth}), $a_\mu^{\mathrm{free}}(a)$ is the corresponding value computed at finite lattice spacing with a physical volume $L=3~\fm$ and $a_\mu^{\mathrm{latt}}(a)$ is the raw lattice data with $L \approx 3~\fm$. 

As for the strange-quark contribution, we have computed the functional derivative with respect to the lattice spacing. In this case we find $\frac{1}{a_\mu(a)} \frac{\partial a_\mu(a)}{\partial a} = 1.013$ very close to one  (so the uncertainty from the scale essentially originates from the explicit muon mass factor in the master equation). Again, this uncertainty from the scale is added in quadrature to the statical uncertainty.
 
\begin{table}[b]
\renewcommand{\arraystretch}{1.25}
\caption{Continuum extrapolations of the charm-quark contributions in units of $10^{-11}$. A ``-'' means that the associated fit parameter is set to zero. The parameters $\beta_2$ and $\gamma$ are always included in the fits. Ensembles with lattice spacings larger than the cut are excluded from the fit. The last line is our final estimate obtained from a model averaging (see~\Table{tab:fitS}). \label{tab:fitC}}
\begin{center}
\begin{tabular}{c@{\quad}c@{\quad}c@{\quad}c@{\quad}|@{\quad}c@{\quad}c@{\quad}|@{\quad}c@{\quad}c@{\quad}|@{\quad}c@{\quad}c}
\hline
 & & & & \multicolumn{2}{c@{\quad}|@{\quad}}{ $\vec{n} = (1,1,1,1)$ } & \multicolumn{2}{c@{\quad}|@{\quad}}{  $\vec{n} = (0,1,1,1)$ } & \multicolumn{2}{c}{  $\vec{n} = (1,1,1,1)$ } \\
\hline
$\beta_3$ & $\beta_4$ & $\delta_2$ & 	Cut & $a_{\mu}^{\conn,c}$ & $\chi^2/\dof$ &	$a_{\mu}^{\conn,c}$ &	$\chi^2/\dof$ &	$a_{\mu}^{\rm (2+2),c}$ &	$\chi^2/\dof$ \\
\hline
 -	&	-	& -		&  -			&	3.918(4)	& 0.26	&	3.835(6) 	&	1.62	&	$-0.189(44)$	& 0.43 	\\
-	&	 -	& -		&  0.12~fm	&	3.923(6)	& 0.14 	&	3.847(8)   	& 	0.71	&	$-0.181(75)$	& 1.25 	\\
-	&	 -	& -		&  0.10~fm	&	3.921(8) 	& 0.07	&	3.848(8) 	&	0.85 	\\ 
-	&	yes	& -		&  -			& 	3.927(10)	& 0.20	&	3.865(14) 	&	0.85 \\
-	&	 -	& $n=2$	&  -			& 	3.927(13) 	& 0.20 	&	3.867(13)	&	0.84	\\
-	&	 -	& $n=3$	&  -			& 	3.926(11)	& 0.19	&	3.863(15)	& 	0.82 	\\
yes	&	yes	& -		&  -			& 	3.868(65)	& 0.10	&	3.786(79)	&	0.80	 \\
\hline
\multicolumn{4}{c@{\quad}|@{\quad}}{ Model averaging} & \multicolumn{2}{c@{\quad}|@{\quad}}{  $3.916(12)_{\stat}(20)_{\syst}$  } & \multicolumn{2}{c@{\quad}|@{\quad}}{ $3.844(17)_{\stat}(26)_{\syst}$ }	& \multicolumn{2}{c@{\quad}}{ $-0.185(52)_{\stat}(4)_{\syst}$ }  \\
\hline
\end{tabular}
\end{center}
\end{table}

\begin{figure}[t]
	\includegraphics*[width=0.49\linewidth]{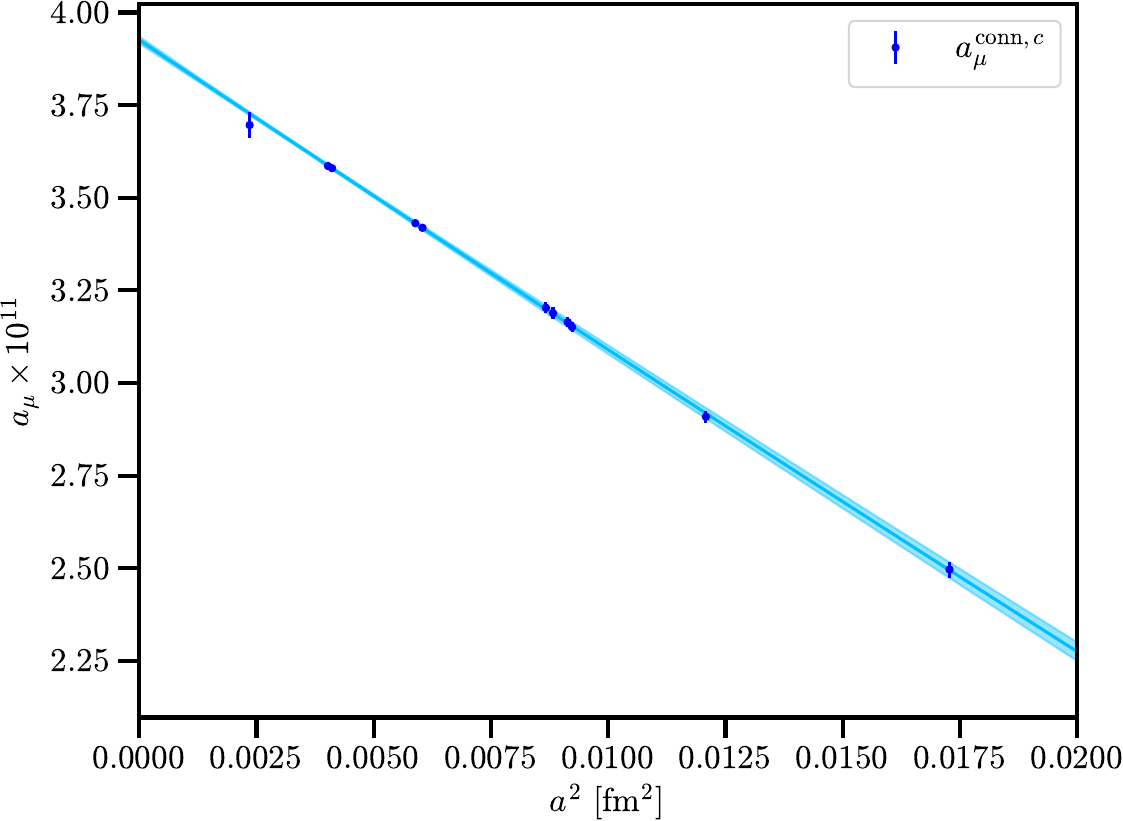}
	\includegraphics*[width=0.49\linewidth]{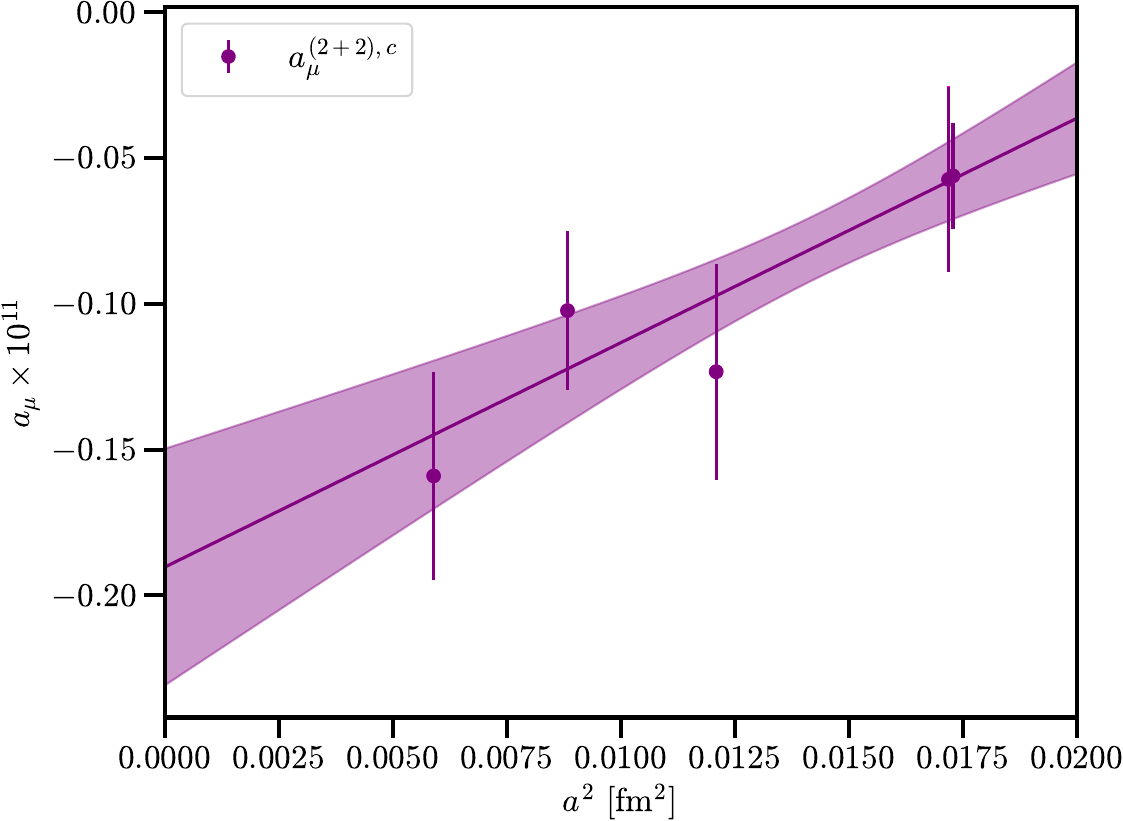}
	\caption{Left: continuum extrapolation of the connected charm-quark contribution with $\vec{n} = (1,1,1,1)$. Right: continuum extrapolation of the leading (2+2) charm-quark disconnected contribution. }
\label{fig:extrapC}
\end{figure}

The charm-quark connected contribution has been computed at five values of the lattice spacing with high statistics. 
The results are summarized in Table~\ref{tab:dataC} and the integrand at our finest lattice spacing is shown in the left panel of~\Fig{fig:integrandC}.   
For each trajectory, we extrapolate our data to the continuum limit assuming the ansatz  
\begin{equation}
a_{\mu}^{\conn, c}(a,M_{\eta_c}) = a_{\mu}^{\conn, c} + \sum_{n=2}^4 \beta_n (\Lambda a)^n + \delta_2 (\Lambda a)^2 \, \alpha_s^n(a^{-1})  + \gamma \frac{\delta M_{\eta_c}}{M_{\eta_c}^\phys} 
\label{eq:fitC}
\end{equation}
where the last term is our proxy for the deviation of the charm quark mass from its physical value.  
Here $M_{\eta_c}$ is the mass obtained from the connected pseudoscalar two-point correlation function and $\delta M_{\eta_c} = M_{\eta_c} - M_{\eta_c}^{\phys}$\ with the experimental value $M_{\eta_c}^{\phys} = 2983.9(4)~\mathrm{MeV}$ taken from~\cite{ParticleDataGroup:2022pth}. 
As for the strange-quark contribution, we have performed several fits, listed in \Table{tab:fitC}, to estimate the systematic uncertainty.  
As can be seen in~\Fig{fig:extrapC}, we observe a close-to-linear scaling in $a^2$, which is supported by the additional ensemble at $a \approx 0.048~\fm$ and large volume $L = 5.5~\fm$, but smaller statistics. However, we note that higher order discretization effects are needed to find reasonable agreement between both trajectories that are extrapolated independently, emphasizing the difficulty of the continuum extrapolation. 
Thus we use a weighted average over both trajectories and use the difference as an additional systematic associated to the continuum extrapolation. This leads to  our final result
\begin{equation}
a_{\mu}^{\conn,c} = 3.916(12)_{\stat}(20)_{\syst}(72)_{\mathrm{cont}}(250)_{\mathrm{kernel}} \times 10^{-11} \,.
\label{res:c}
\end{equation}
The last systematic error, associated to the QED weight function, is discussed below. The order of magnitude is similar to the connected strange-quark contribution. This is partly explained by the larger charge factor of the charm quark compared to the strange quark that compensates the heavier quark mass.

\subsection{Further checks}

\begin{figure}[t]
	\includegraphics*[width=0.48\linewidth]{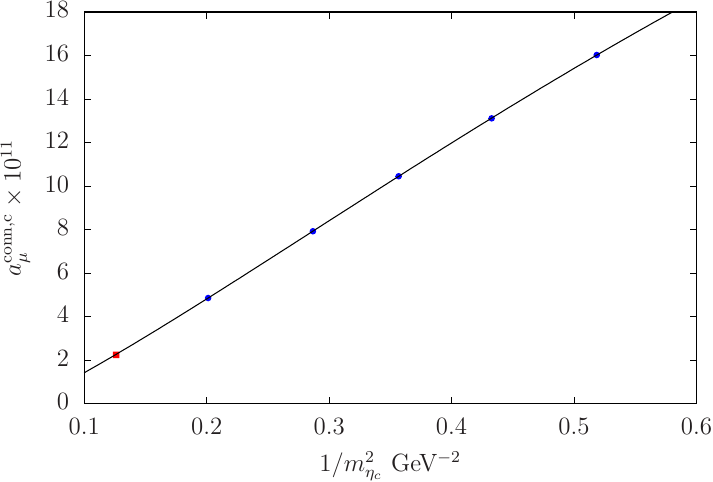}
	\includegraphics*[width=0.49\linewidth]{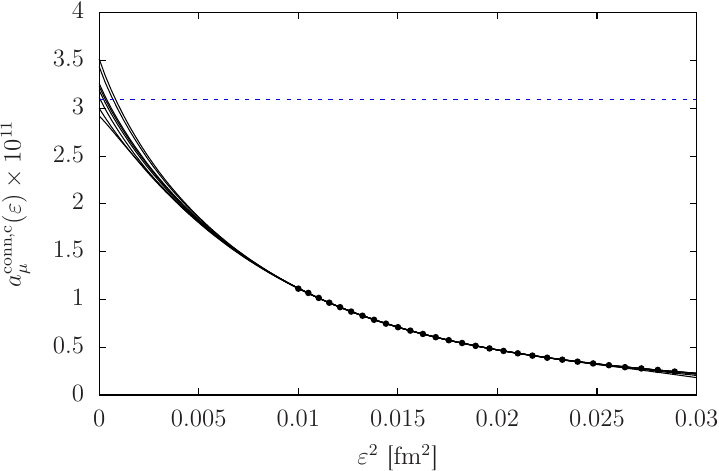}	
	\caption{Left: Extrapolation of $a_{\mu}^{\rm conn,c}$ as a function of $1/m_{\eta_c}^2$ at the lattice spacing $a \approx 0.095~\fm$. The red square is the simulation performed at the physical charm quark mass while the blue circles are obtained using lighter than physical charm quark masses. The line represents our extrapolation excluding the red point. 
	Right: extrapolation of the lepton-loop contribution (computed in the continuum limit), as a function of the regulator $\varepsilon$. The results correspond to $m_{\rm lepton}/m_{\mu} = 12.02$. The horizontal dashed blue line is the known exact value.}
\label{fig:charmcheck}
\end{figure}

Smaller than physical charm quark masses allow to better sample the integrand at the cost of an additional extrapolation towards the physical charm mass. This is the strategy followed in~\cite{Chao:2022xzg}. 
Thus, at a single value of the lattice spacing, $a \approx 0.095~\fm$, we perform a dedicated study of the quark mass dependence of $a_{\mu}^{\rm conn,c}$. This observable is computed at six values of the (valence) charm mass such that the mass of the pseudoscalar $\eta_c$ meson lies in the range [1.39:2.82]~GeV (this range is chosen to approximatively match the values used in~\cite{Chao:2022xzg}). 
At large quark masses, we expect $a_\mu$ to scale as $m_{\mu}^2 / m_{\eta_c}^2$. 
The data are extrapolated using a polynomial of order 3 in $1/m_{\eta_c}^2$ where a constant term, that can be present at finite lattice spacing, is allowed. 
The result of the extrapolation is shown in the left panel of \Fig{fig:charmcheck}. Excluding the point at the physical charm-quark mass the result of the extrapolation $2.23(5) \times 10^{-11}$ is in excellent agreement 
(at the level of precision quoted in~\Eq{res:c}) 
with the values $2.2525(9) \times 10^{-11}$ obtained by the direct simulation at the physical charm quark mass. The quality of the fit, $\chi^2 / \dof = 0.97$, is very good. 

In~\cite{Chao:2022xzg}, it was noticed that the numerical evaluation of the QED weight function in~\Eq{eq:lamsub} is challenging at very short distances when either $|x| \to 0$ or $|x-y| \to 0$. However, we note that these numerical instabilities become relevant at distances smaller than those reached in our QCD simulations (this would not be the case with the trajectories $\vec{n} = (0,0,1,1)$ or $\vec{n} = (0,0,0,1)$).
Nevertheless, to estimate possible effects associated with those instabilities, we compute the lepton-loop contribution in the continuum limit using the method described in~\cite{Asmussen:2022oql} for the ratio $m_{\rm lepton}/m_\mu=12.02$ corresponding to $m_{\rm lepton} = m_c$. 
To stabilize the integration, we introduce a regulator (a smooth step function
that varies from 0 to 1 in the range $[\varepsilon/2, \varepsilon]$) to supress points such that $|x| < \varepsilon$ or $|x-y| < \varepsilon$. 
The regulated observable $a_{\mu}^{\rm lepton}(\varepsilon)$ is then computed for several values of $\varepsilon$ in the range [0.1:0.17]~fm. i.e. including distances smaller than those reached in our QCD simulations. We finally extrapolate the result to $\varepsilon \to 0$ assuming a polynomial dependence and including terms up to $ \alpha_6 \varepsilon^6$. Several variations, which include or not the $\alpha_5$ and $\alpha_6$ terms or by performing cuts in $\varepsilon$, are performed. The spread over all variations is shown in the right panel of~\Fig{fig:charmcheck} with a comparison with the known result $a_{\mu}^{\rm lepton} \approx 3.12 \times 10^{-11}$~\cite{Laporta:1992pa}. We take half of the spread as an additional systematic uncertainty associated to the weight function (third uncertainty in~\Eq{res:c}).

\subsection{Leading quark-disconnected contributions}

The leading quark-disconnected contribution is computed with high statistics on five ensembles for the trajectory $\vec{n}  = (1, 1, 1, 1)$ and the results are listed in~\Table{tab:dataC}. The integrands and the partial sums are displayed in~Fig.~\ref{fig:integrandC}. 
First, we note that the charm-light contribution is dominant compared to the charm-strange and charm-charm contributions. 
Second, the disconnected contribution is small compared to the connected contribution but also with respect to the target precision on $\ahlbl$. We extrapolate the sum over all flavor combinations assuming the functional dependence~in \Eq{eq:fitC} and setting $\gamma = \delta_2 = \beta_3 = \beta_4 = 0$. The extrapolation is shown in the right panel of~\Fig{fig:extrapC} and we quote as our final result
\begin{equation}
a_{\mu}^{\rm (2+2),c} = -0.185(52)_\stat(4)_\syst \times 10^{-11} \,,
\end{equation}
where the first error is statistical and the second is the systematic uncertainty associated with the continuum extrapolation obtained by excluding the coarsest lattice spacing from the fit.

\section{sub-leading disconnected contributions \label{sec:sub}} 

In this section, we discuss the sub-leading contributions originating from diagrams with at least one closed vector loop. These contributions vanish exactly in the SU(3) flavor-symmetric limit. Upper bounds on these contributions have been given in~\cite{Chao:2021tvp,Blum:2019ugy}.

\subsection{Master formulae}

\begin{figure}[t]
	\includegraphics*[width=0.85\linewidth]{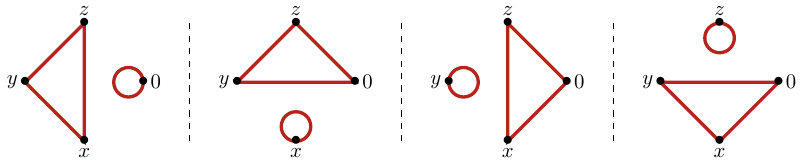}
	\caption{The four contributions to the $(3+1)$ quark-disconnected contribution. They are labelled $\Pi^{(A)}, \Pi^{(B)}, \Pi^{(C)}$ and $\Pi^{(D)}$ from left to right. Each diagram appears twice with opposite fermionic flow in the triangle part.}
\label{fig:3p1}
\end{figure}

There are four diagrams for the (3+1) disconnected contribution to the four-point function. They are depicted in \Fig{fig:3p1} and we denote the corresponding Wick contractions $\Pi^{(A,f)}, \Pi^{(B,f)}, \Pi^{(C,f)}$ and $\Pi^{(D,f)}$, with $f$ the flavor index of the quark in the triangle function. 
Using translational invariance, it can be shown that
\begin{subequations}
\begin{align}
\Pi^{(A,f)}_{\mu\nu\sigma\lambda}(x,y,z) &= \Pi^{(B,f)}_{\lambda\nu\sigma\mu}(-x,y-x,z-x) \,, \\
\Pi^{(C,f)}_{\mu\nu\sigma\lambda}(x,y,z) &= \Pi^{(B,f)}_{\nu\mu\sigma\lambda}(y,x,z) \,,
\end{align}
\end{subequations}
with
\begin{subequations}
\begin{align}
\Pi^{(B,f)}_{\mu\nu\sigma\lambda}(x,y,z) &= T^{(f)}_{\nu\sigma\lambda}(y,z,0) \, L_\mu(x) \,, \\
\Pi^{(D,f)}_{\mu\nu\sigma\lambda}(x,y,z) &= T^{(f)}_{\mu\nu\lambda}(x,y,0) \, L_\sigma(z) \,,
\end{align}
\end{subequations}
and where we have defined the loop and triangle functions by
\begin{subequations}
\begin{align}
L_\mu(x) &= - \frac{1}{3} \, \mathrm{Im} \, \Tr\left[ G_l(x,x) \gamma_{\mu} - G_s(x,x) \gamma_{\mu} \right] \,, \label{eq:loop} \\
T^{(f)}_{\mu\nu\lambda}(x,y,z) &= - 2 \, \mathrm{Im} \, \Tr\left[ \gamma_\mu G_f(x,y) \gamma_\nu G_f(y,z) \gamma_\lambda G_f(z,x) \right]  \,.
\end{align}
\end{subequations}
Only the light and strange-quark contributions are taken into account when evaluating the loop functions. We assume isospin symmetry ($m_u=m_d$) and the electric charge factor $\mathcal{Q}_u + \mathcal{Q}_d - \mathcal{Q}_s=1/3$ is included in the definition of the loop functions.
Thus, an estimator for the (3+1) quark-disconnected contribution is given by
\begin{multline}
\mathcal{I}^{(3+1)}(|y|) =  \frac{m_{\mu}e^6}{3} \sum_{f=u,d} \mathcal{Q}_f^3 \, 2\pi^2 |y|^3 \times \\ \sum_{x,z} \Lsym(x,z) \left[ (3 z_{\rho}  - x_{\rho} ) T^{(f)}_{\nu\sigma\lambda}(y,z,0) \, L_\mu(x) +  z_{\rho} T^{(f)}_{\mu\nu\lambda}(x,y,0) \, L_\sigma(z) \right] \,.
\end{multline}
In practice, we restrict the sum over $f$ to the light $u$ and $d$ quarks and we ignore the strange-quark contribution in the triangle function. This approximation is motivated by the fact that this contribution is further suppressed by the ratio of charge factors $-1/7$.

\begin{figure}[t]
	\includegraphics*[width=0.99\linewidth]{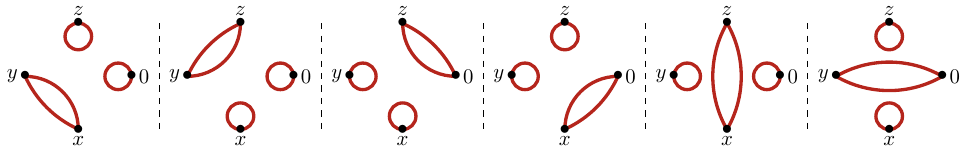}
	\caption{The six contributions to the $(2+1+1)$ quark-disconnected contribution. Each diagram appears twice with opposite fermionic flow in the two-point function.}
\label{fig:2p1p1}
\end{figure}

The (2+1+1) contribution can be estimated from the subtracted two-point functions defined in \Eq{eq:pihat} and the loop function of \Eq{eq:loop}. In that case, we have six contributions depicted in \Fig{fig:2p1p1}. 
By exchanging $(x,\mu) \leftrightarrow (y,\nu)$ and using translational invariance, the fifth and sixth diagrams can be expressed in terms of the second and fourth diagrams. Thus our final estimator reads 
\begin{multline}
\mathcal{I}^{(2+1+1)}(|y|) =  \frac{m_{\mu}e^6}{3} \sum_{f=u,d} \mathcal{Q}_f^2 \, 2\pi^2 |y|^3 \sum_{x,z} \Lsym(x,y) \, z_{\rho} \, \left[  
\widehat{\Pi}^{(f)}_{\mu\nu}(x,y) \, L_\sigma(z) \, L_\lambda(0)  + \right. \\ \left.
2 \, \widehat{\Pi}^{(f)}_{\nu\sigma}(y,z) \, L_\mu(x) \, L_\lambda(0) +
\widehat{\Pi}^{(f)}_{\sigma\lambda}(z,0) \, L_\mu(x) \, L_\nu(y) +
2 \, \widehat{\Pi}^{(f)}_{\mu\lambda}(x,0) \, L_\nu(y) \, L_\sigma(z) 
\right] \,.
\end{multline}

Finally, for the (1+1+1+1) quark-disconnected contribution, the four-point correlation function is expressed as a product of four loop functions where all non-vanishing vacuum expectation values have been subtracted out
\begin{align}
\nonumber \mathcal{I}^{(1+1+1+1)}(|y|) = - \frac{m_{\mu}e^6}{3} \, 2\pi^2 |y|^3\, \sum_{x,z} \Lsym(x,y) z_{\rho} \, \Big[ & \langle L_{\mu}(x) L_{\nu}(y) L_{\sigma}(z) L_{\lambda}(0)  \rangle_U  \\[-4mm]
\nonumber - & \langle L_{\mu}(x) L_{\nu}(y) \rangle_U \, \langle L_{\sigma}(z) L_{\lambda}(0)  \rangle_U \\
\nonumber - & \langle L_{\mu}(x) L_{\sigma}(z) \rangle_U \, \langle L_{\nu}(y) L_{\lambda}(0)  \rangle_U \\
 - &          \langle L_{\nu}(y) L_{\sigma}(z) \rangle_U \, \langle L_{\mu}(x) L_{\lambda}(0)  \rangle_U \Big] \,.
\end{align}

\subsection{Lattice details}

These contributions are computed using the same method as the one described for the connected and leading quark-disconnected contributions. 
The loop functions are evaluated using the noise reduction techniques described in~\cite{spectro}. 
It includes low-mode averaging~\cite{DeGrand:2004wh,Giusti:2004yp} and the point-split estimator introduced in~\cite{Giusti:2019kff}. We analyze	 three ensembles at three values of the lattice spacing. See \Table{tab:ens2}.

\begin{table}[b]
\renewcommand{\arraystretch}{1.1}
\caption{Same as \Table{tab:ens} for the sub-leading quark-disconnected contributions.}
\vskip 0.1in
\begin{tabular}{l@{\hskip 01em}c@{\hskip 01em}l@{\hskip 01em}c@{\hskip 01em}l@{\hskip 01em}l@{\hskip 01em}l@{\hskip 01em}l}
\hline
$\quad\beta\quad$	&	$L^3\times T$ 	&	$a~[\fm]$	&	$L~[\fm]$	&	$am_l$		&	$am_s$	&	$\#$confs	 	 & Id\\
\hline
$3.7000$	& $24^3\times48$ & 0.132 & 3.2 & 0.00205349 & 0.0572911 & 700 & V3-L24\\
\hline
$3.7553$	& $28^3\times56$ & 0.112 & 3.1 & 0.00171008 & 0.0476146 & 850 & V6-L28\\
\hline
$3.8400$	& $32^3\times64$ & 0.095 & 3.0 & 0.001455 & 0.03913 & 1000	 & V3-L32-4\\
\hline
 \end{tabular} 
\label{tab:ens2}
\caption{Value of the (3+1) disconnected contribution on each ensemble. Lattice data are used for $|y| \leq \ycut$. We add a 100\% systematic uncertainty to the tail contribution.}
\begin{center}
\begin{tabular}{l@{\quad}c@{\quad}c}
\hline
Id		& $\ycut$~[fm] & $a_{\mu}^{(3+1),l} \times 10^{11}$ \\ 
\hline
V3-L24	& 1.84 & $0.80(37)_\stat(33)_\syst$ \\
\hline
V3-L28	& 1.98 & $0.68(25)_\stat(19)_\syst$ \\
\hline
V3-L32-4	& 1.88 & $1.39(31)_\stat(54)_\syst$ \\
\hline 
\end{tabular}
\label{tab:3p1}
\end{center}
\end{table}

\subsection{Results}

\begin{figure}[t]
	\includegraphics*[width=0.32\linewidth]{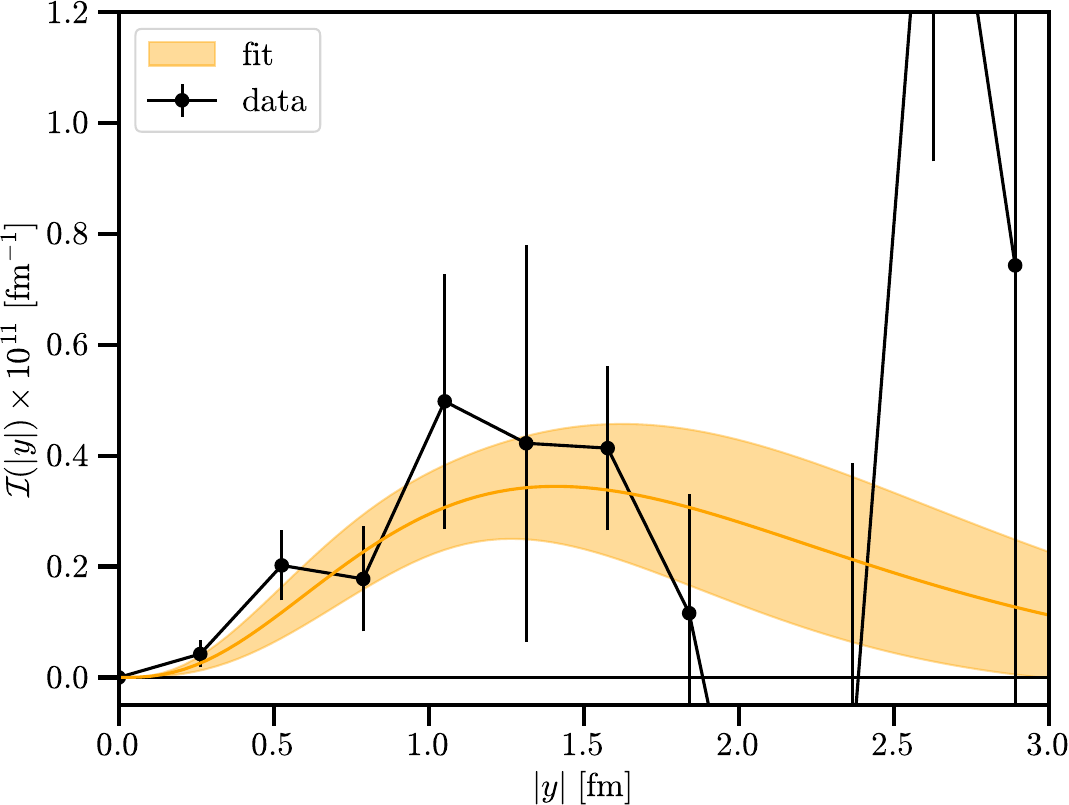}
	\includegraphics*[width=0.32\linewidth]{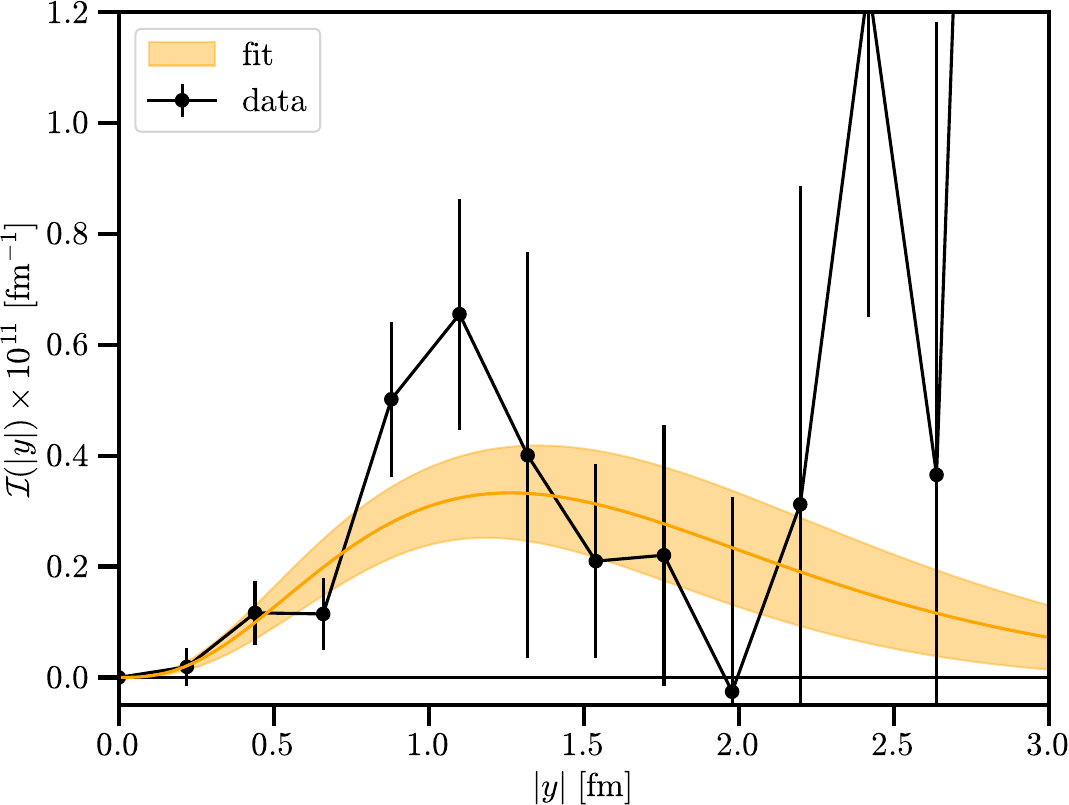}
	\includegraphics*[width=0.32\linewidth]{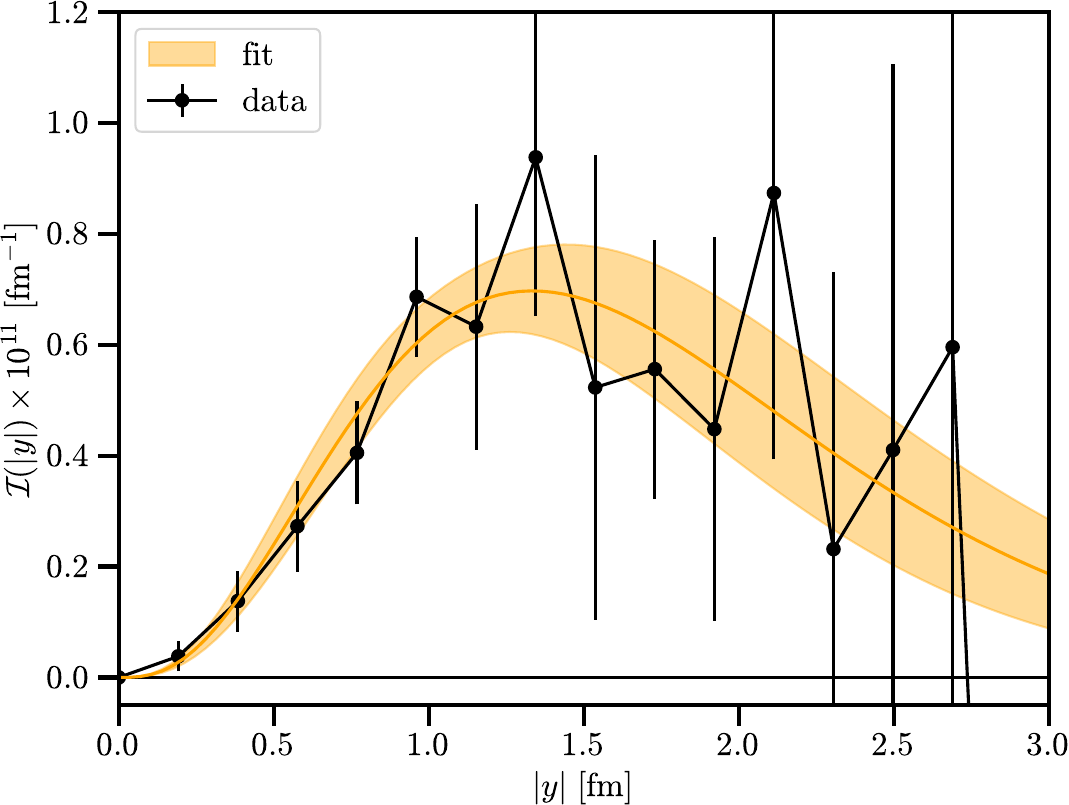}
	\caption{Integrand for the (3+1) quark-disconnected contribution at our three values of the lattice spacings (the coarsest is on the left, the finest on the right). The bands are the results of the fit described in the main text and used to estimate the tail with $y>\ycut$.}
\label{fig:Isub}
\end{figure}

The integrand for the (3+1) quark-disconnected contribution is shown in \Fig{fig:Isub} at three values of the lattice spacing. A clear signal is observed at short distances and the signal is eventually lost at $|y| \approx 2~\fm$. Thus, we integrate the lattice data up to $\ycut \approx 2~\fm$. To estimate the missing tail contribution, we assume the functional dependence $\mathcal{I}(z) = A |y|^3 \exp(-B |y|)$ where $A$ and $B$ are free fit parameters and we associate 100\% uncertainty to this tail contribution. The results are summarized in~\Table{tab:3p1}. 
Because of the large uncertainties, a constant continuum extrapolation suffices, leading to our final estimate
\begin{equation}
a_{\mu}^{(3+1)} = 0.83(25) \times 10^{-11} \,.
\end{equation}
This contribution is well below the target precision of 10\% on $\ahlbl$.

\begin{figure}[t]
	\includegraphics*[width=0.41\linewidth]{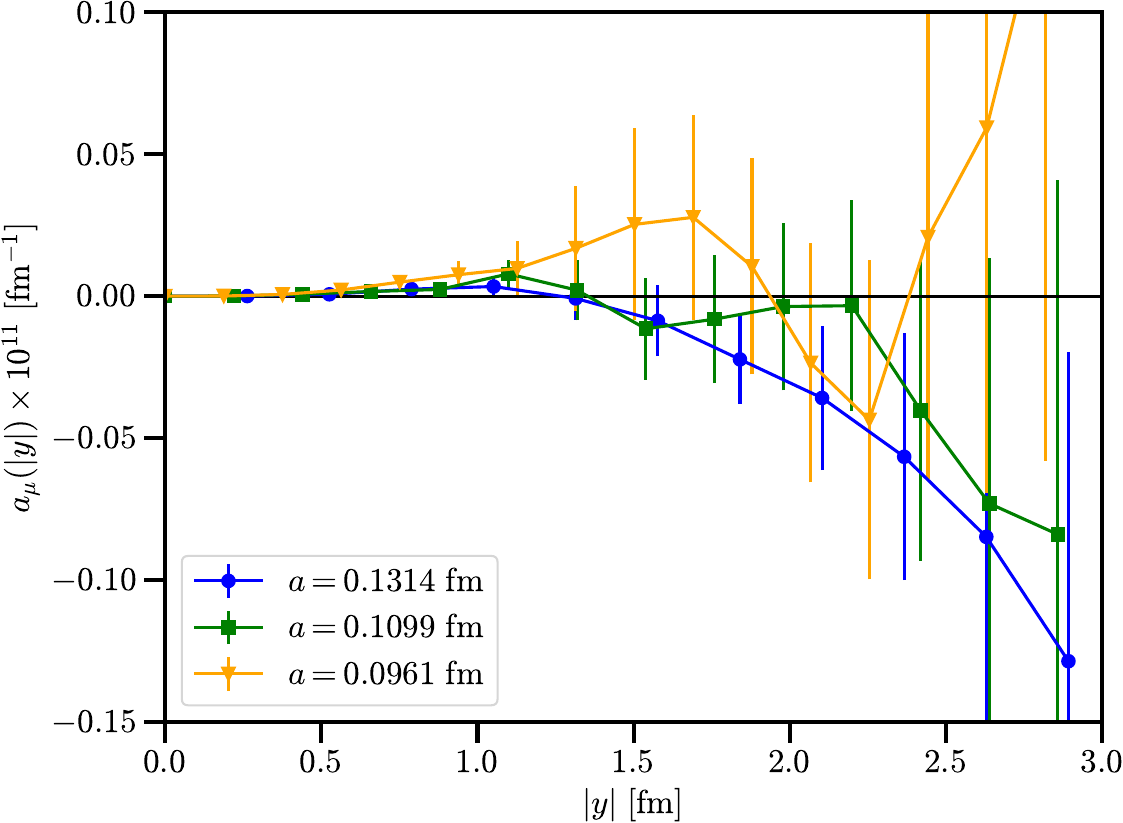} \qquad
	\includegraphics*[width=0.41\linewidth]{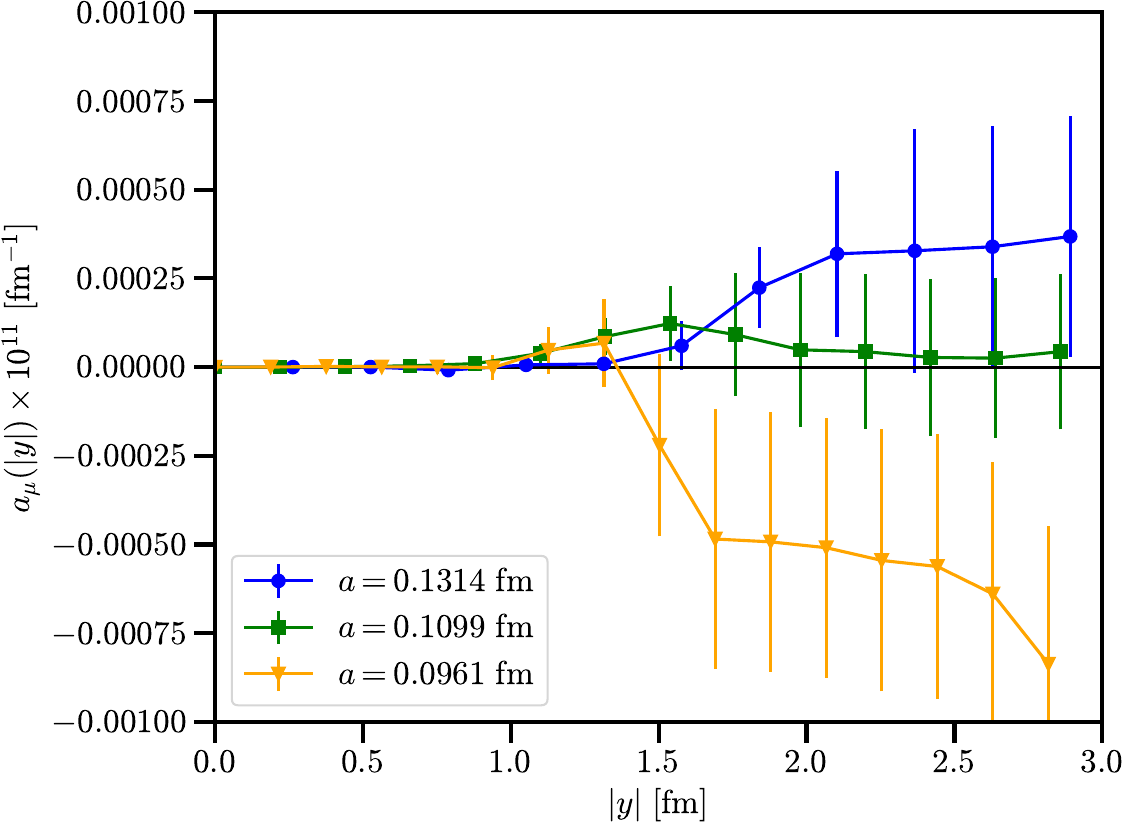}
	\caption{Partial sums for the (2+1+1) contribution (left) and for the (1+1+1+1) contribution (right).}
\label{fig:asub}
\end{figure}

For the (2+1+1) and (1+1+1+1) contributions, the partial sums are shown in~\Fig{fig:asub}. The statistical precision is too low and the signal is compatible with zero. 
Since the (3+1) strange-quark contribution is anyway neglected, we simply provide upper bounds on the absolute value of the contributions. 
In~\Fig{fig:asub}, no clear dependence on the lattice spacing is observed and we use the largest value of $a_{\mu}$ among all three ensembles, at $\ycut \approx 2~\fm$, as a bound
\begin{subequations}
\begin{align}
|a_{\mu}^{(2+1)}| &\lessapprox 0.04 \times 10^{-11} \,, \\
|a_{\mu}^{(1+1+1+1)}| &\lessapprox 5 \times 10^{-15} \,.
\end{align}
\end{subequations}

\section{Conclusion \label{sec:ccl}}  

\begin{table}[b]
\caption{Summary of all individual contributions to $\ahlbl$ at the physical point and in the continuum limit. The final result is obtained by adding the total light, strange and charm quark contributions with the sub-leading quark-disconnected contribution.}
\begin{center}
\begin{tabular}{l@{\qquad}c}
\hline
Contribution		&	$a_{\mu} \times 10^{11}$ \\
\hline
Light Connected 	&	\phantom{$-$}220.1(13.5) \\
Light 2+2		 	&	$-101.1(12.8)$ \\
Light Total			&	\phantom{$-$}122.6(11.6)  \\
\hline
Strange Connected 	&	$\phantom{-}3.694(25)_{\stat}(8)_{\syst}$ \\ 
Strange 2+2		&	$-5.4(0.8)_\stat(0.2)_\syst$ \\
Strange Total		&	$-1.7(0.8)_\stat(0.3)_\syst$ \\
\hline
Charm Connected 	&	$\phantom{-}3.92(1)_\stat(26)_{\syst}$ \\
Charm 2+2		&	$-0.185(51)_\stat(04)_\syst$ \\
Charm Total		&	$\phantom{-}3.73(5)_\stat(26)_{\syst}$ \\
\hline
sub-leading Disc.	&	$0.83(25)_{\stat}$ \\
\hline
\textbf{Total}		&	$125.5(11.6)_\stat(0.4)_\syst$ \\
\hline
\end{tabular}
\end{center}
\label{tab:final}
\end{table}

\begin{figure}[t]
	\includegraphics*[width=0.65\linewidth]{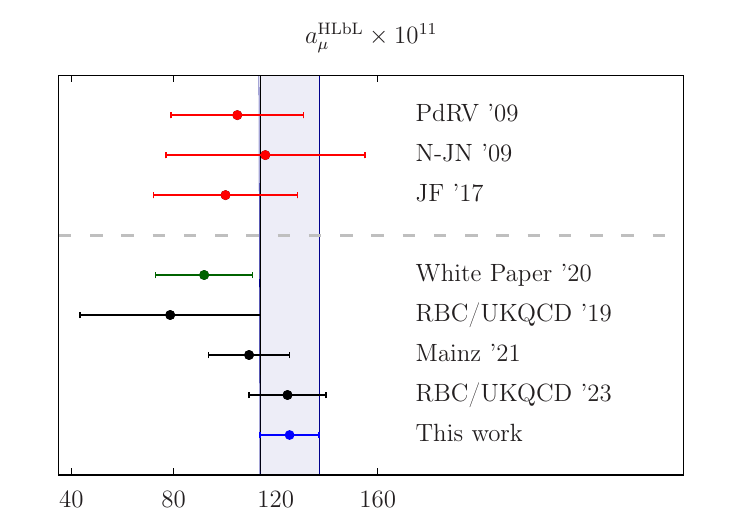}
	\caption{Status of HLbL calculations. In green, the 2020 white paper estimate~\cite{Aoyama:2020ynm}. Lattice calculations from the RBC/UKQCD collaboration~\cite{Blum:2023vlm,Blum:2019ugy,Blum:2017cer} and from the Mainz group~\cite{Chao:2022xzg,Chao:2021tvp,Chao:2020kwq}. In red, the older model-dependent estimates~\cite{Jegerlehner:2017lbd,Prades:2009tw,Nyffeler:2009tw,Jegerlehner:2009ry}.}
\label{fig:status}
\end{figure}

In this work we have presented a complete lattice calculation of the HLbL contribution to the anomalous magnetic moment of the muon using staggered quarks at the physical point. 
A summary of all individual contributions is provided in Table~\ref{tab:final}. The final estimate is obtained by adding the light, strange and charm-quark contributions as well as the sub-leading quark-disconnected contributions. Statistical uncertainties are added in quadrature while systematics, which are partially correlated, are added linearly. Our final value reads
\begin{equation}
\ahlbl = 125.5(11.6)_{\stat}(0.4)_{\syst} \times 10^{-11} \,.
\end{equation}
Our error budget is completely dominated by the statistical uncertainty on the light-quark contribution and the achieved precision is already sufficient in view of the forthcoming measurement by the Fermilab collaboration, where a precision of 0.14 ppm is expected. 
In the future, we plan to add an additional gauge ensemble with a finer lattice spacing and high statistics, in order to further improve our estimate. 
A comparison of our result with previous determinations is shown in \Fig{fig:status}. We observe a good agreement among all lattice calculations and our result lies $1.5$ standard deviations above the data-driven approach estimate of~\cite{Aoyama:2020ynm} based on~\cite{Melnikov:2003xd,Masjuan:2017tvw,Colangelo:2017fiz,Hoferichter:2018kwz,Gerardin:2019vio,Bijnens:2019ghy,Colangelo:2019uex,Pauk:2014rta,Danilkin:2016hnh,Jegerlehner:2017gek,Knecht:2018sci,Eichmann:2019bqf,Roig:2019reh}.

It is also useful to compare individual contributions. These are affected by different systematics. 
For the light-quark contribution, all lattice collaborations quote results for both the connected and leading (2+2) quark-disconnected contributions at the physical point and in the continuum limit. Our result for the connected contribution is smaller than the RBC/UKQCD value by $1.3\sigma$ and than the Mainz result, by $1.4\sigma$. 
For the (2+2) disconnected contribution our result is smaller in magnitude by $1.1\sigma$ and $1.3\sigma$ with RBC/UKQCD and the Mainz group respectively. All lattice results agree within one standard deviation for the total light-quark contribution where both finite-volume effects and the pion-mass dependance are significantly reduced compared to connected and leading quark-disconnected contributions taken individually. 
The total strange-quark contribution is also in good agreement with previous lattice calculations. 
However, for the connected contribution, where a high statistical precision can be reached, our value lies about two sigma above the RBC/UKQCD result~\cite{Blum:2019ugy}. 
Concerning the charm-quark contribution, our result is $1.7\sigma$ larger than the result obtained by the Mainz group~\cite{Chao:2022xzg}. Ours is computed  directly at the physical quark mass. Six lattice spacings are used for the continuum extrapolation and, at the finest lattice spacing, the bare charm quark mass is $am_c \approx 0.24 \ll 1$. 
Nevertheless, at the physical charm quark mass the continuum extrapolation is challenging and currently limited by the precision of the QED weight function (see \Eq{eq:lamsub}) as discussed in~\Section{sec:charm}. 
Probing smaller lattice spacings may require a numerically more stable implementation of the QED weight function at very short distances. 
Because of those challenges the uncertainty of our result is systematics dominated, as is the one of the Mainz estimate. Thus, the significance of tension should be taken with caution.

Finally, a clear signal is observed for the (3+1) quark-disconnected contribution and we confirm the smallness of all the sub-leading disconnected contributions compared to the target precision. 

\begin{acknowledgments}
We thank all the members of the Budapest-Marseille-Wuppertal collaboration for helpful discussions and the access to the gauge ensembles used in this work. 
This publication received funding from the Excellence Initiative of Aix-Marseille University - A*Midex, a French ``Investissements d'Avenir" programme, AMX-18-ACE-005 and from the French National Research Agency under the contract ANR-20-CE31-0016. 
This project was provided with computer and storage resources by GENCI on the supercomputers Adastra at CINES, Jean Zay at IDRIS and Joliot-Curie at CEA's  TGCC via the grants 0511504 and 0502275. 
The authors gratefully acknowledge the Gauss Centre for Supercomputing (GCS) e.V. for providing computer time on the GCS supercomputers SuperMUC-NG at Leibniz Supercomputing Centre in M\"{u}nchen, HAWK at the High Performance Computing Center in Stuttgart, JUWELS and JURECA at Forschungszentrum J\"{u}lich.
Centre de Calcul Intensif d'Aix-Marseille (CCIAM) is acknowledged for granting access to its high-performance computing resources.\\
\end{acknowledgments}

\bibliography{biblio}{}

\end{document}